\documentclass[article,shortnames,nojss]{jss} 
\pdfoutput=1
\usepackage{amsmath}
\usepackage{amsfonts}
\usepackage{amssymb}
\usepackage{graphicx}
\usepackage{algorithm}
\usepackage[latin1]{inputenc}
\usepackage[T1]{fontenc}
\usepackage{fancyhdr}
\usepackage{subfigure}
\usepackage{epstopdf}
\usepackage{hyperref}
\usepackage{fancybox}
\usepackage{tikz}
\usepackage{listings}
\usepackage{afterpage}

\usepackage{lmodern}
\usepackage{amsthm}
\theoremstyle{plain}

\usetikzlibrary{arrows}

\newtheorem*{example2}{Example}

\newcommand{\Biips}{\pkg{Biips}}
\newcommand{\Rbiips}{\pkg{Rbiips}}
\newcommand{\Matbiips}{\pkg{Matbiips}}
\newcommand{\JAGS}{\pkg{JAGS}}
\newcommand{\rjags}{\pkg{rjags}}
\newcommand{\BUGS}{\proglang{BUGS}}
\newcommand{\OpenBUGS}{\pkg{OpenBUGS}}
\newcommand{\WinBUGS}{\pkg{WinBUGS}}
\newcommand{\Stan}{\pkg{Stan}}
\newcommand{\R}{\proglang{R}}
\newcommand{\Cpp}{\proglang{C++}}
\newcommand{\Matlab}{\proglang{MATLAB}}

\newcommand{\Octave}{\proglang{Octave}}
\newcommand{\SMCTC}{\pkg{SMCTC}}
\newcommand{\LibBi}{\pkg{LibBi}}

\DeclareMathOperator{\Norm}{\mathcal N}
\DeclareMathOperator{\TNorm}{\mathcal{TN}}

\DeclareMathOperator*{\Discrete}{Discrete}

\DeclareMathOperator{\Gam}{Gamma}
\DeclareMathOperator{\Beta}{Beta}

\DeclareMathOperator{\SESS}{SESS}

\graphicspath{{./}{./figures/}}

\definecolor{mygreen}{rgb}{0,0.6,0}
\definecolor{mygray}{rgb}{0.5,0.5,0.5}
\definecolor{mymauve}{rgb}{0.58,0,0.82}
\definecolor{shadecolor}{rgb}{0.9,0.9,0.9}
\definecolor{shadecolor2}{rgb}{.95,1,0.95}
\definecolor{shadecolor3}{rgb}{.95,0.95,1}
\definecolor{darkblue}{rgb}{0,0,.5}
\definecolor{darkgreen}{rgb}{0,0.5,0}
\definecolor{lightred}{rgb}{1.0,0.95,0.92}

\lstdefinestyle{matbiips}{
  backgroundcolor=\color{lightred},   
  basicstyle=\footnotesize\ttfamily,        
  breakatwhitespace=false,         
  breaklines=true,                 
  captionpos=t,                    
  commentstyle=\color{mygreen},    
  escapeinside={\%*}{*)},          
  extendedchars=true,              
  frame=none,                    
  keepspaces=true,                 
  keywordstyle=\color{blue},       
  keywordstyle=[2]\color{blue}\bfseries,
  language=Matlab,                 
  deletekeywords={mean,pi,diag},            
  morekeywords={},            
  morekeywords=[2]{biips_smc_sensitivity,biips_pmmh_update,
  biips_pmmh_samples,biips_init,biips_pimh_init,biips_pmmh_init,
  biips_smc_samples,biips_diagnosis,
  biips_pimh_update,biips_summary,biips_pimh_samples,
  biips_model,biips_density,biips_add_function,biips_add_distribution},
  numbers=none,                    
  numbersep=5pt,                   
  numberstyle=\tiny\color{mygray}, 
  rulecolor=\color{black},         
  showspaces=false,                
  showstringspaces=false,          
  showtabs=false,                  
  stepnumber=2,                    
  stringstyle=\color{mymauve},     
  tabsize=2,                       
  title={\footnotesize Matbiips},                   
}

\lstdefinestyle{Rbiips}{
  language=R,                
  basicstyle=\footnotesize\ttfamily,           
  numbers=none,                   
  numberstyle=\tiny\color{gray},  
  stepnumber=2,                   
  numbersep=5pt,                  
  backgroundcolor=\color{shadecolor2},      
  showspaces=false,               
  showstringspaces=false,         
  showtabs=false,                 
  frame=none,                   
  rulecolor=\color{black},        
  tabsize=2,                      
  captionpos=t,                   
  breaklines=true,                
  breakatwhitespace=false,        
  title={\footnotesize Rbiips},                   
  keywordstyle=\color{blue},          
  keywordstyle=[2]\color{blue}\bfseries,
  commentstyle=\color{darkgreen},       
  stringstyle=\color{mymauve},         
  escapeinside={\%*}{*)},            
  deletekeywords={which,dt,sum,log,runif,cumsum,dim,names,range,seq,rep,length,nrow,list,t_max,t,mean,pi,diag,matrix,c, max,_,<-,model,file,sample,data},
  otherkeywords={biips_smc_sensitivity,biips_pmmh_update,
  biips_pmmh_samples,biips_init,biips_pimh_init,biips_pmmh_init,
  biips_smc_samples,biips_diagnosis,
  biips_pimh_update,biips_summary,biips_pimh_samples,
  biips_model,biips_density,biips_add_function,biips_add_distribution},
  morekeywords={}      ,         
  morekeywords=[2]{biips_smc_sensitivity,biips_pmmh_update,
  biips_pmmh_samples,biips_init,biips_pimh_init,biips_pmmh_init,
  biips_smc_samples,biips_diagnosis,
  biips_pimh_update,biips_summary,biips_pimh_samples,
  biips_model,biips_density,biips_add_function,biips_add_distribution},
  stringstyle=\color{mymauve},     
  }

\lstdefinestyle{bugs}{
  basicstyle=\footnotesize\ttfamily,           
  numbers=none,                   
  numberstyle=\tiny\color{gray},  
  stepnumber=2,                   
  numbersep=5pt,                  
  backgroundcolor=\color{shadecolor3},      
  showspaces=false,               
  showstringspaces=false,         
  showtabs=false,                 
  frame=none,                   
  rulecolor=\color{black},        
  tabsize=2,                      
  captionpos=t,                   
  breaklines=true,                
  breakatwhitespace=false,        
  title={\tiny \lstname },                   
  keywordstyle=\color{blue},          
  keywordstyle=[2]\color{blue}\bfseries,          
  commentstyle=\color{darkgreen},       
  stringstyle=\color{mymauve},         
  escapeinside={\%*}{*)},            
  morekeywords={dnorm,dgamma,dcat,for, <-},               
  morekeywords=[2]{model, data,var}               
  }

\tikzstyle{every node}=[font=\footnotesize]

\tikzstyle{state}=[circle,
                                    thick,
                                    draw=blue!80,
                                    fill=blue!20
                                    ]

\tikzstyle{measurement}=[circle,
                                                thick,
                                                draw=orange!80,
                                                fill=orange!25]
\tikzstyle{param}=[circle,
                                                thick,
                                                draw=green!80,
                                                fill=green!25]

\tikzstyle{input}=[rectangle, rounded corners,
                                    thick,
                                    text width=3cm,
                                    draw=blue!80,
                                    fill=blue!20,
                                    text centered,
                                    font=\large,
                                    ]
\tikzstyle{processing}=[rectangle, rounded corners,
                                    thick,
                                    text width=3cm,
                                    draw=green!80,
                                    fill=green!20,
                                    text centered,
                                    font=\large,
                                    ]
\tikzstyle{output}=[rectangle, rounded corners,
                                                thick,
                                                text width=3cm,
                                                draw=orange!80,
                                                fill=orange!25,
                                                text centered,
                                    font=\large,]

\author{Adrien Todeschini\\INRIA\And
        Fran\c cois Caron\\University of Oxford\And
        Marc Fuentes\\INRIA\AND
        Pierrick Legrand\\University of Bordeaux\And
        Pierre Del Moral\\University of New South Wales
        }
\title{\Biips: Software for Bayesian Inference with Interacting Particle Systems}

\Plainauthor{A. Todeschini, F. Caron, M. Fuentes, P. Legrand, P. Del Moral} 
\Plaintitle{Biips: a Software for Bayesian Inference with Interacting Particle Systems} 
\Shorttitle{\Biips} 

\Abstract{
 \Biips\ is a software platform for automatic Bayesian inference with interacting particle systems. \Biips\ allows users to define their statistical model in the probabilistic programming \BUGS\ language, as well as to add custom functions or samplers within this language. Then it runs sequential Monte Carlo based algorithms (particle filters, particle independent Metropolis-Hastings, particle marginal Metropolis-Hastings) in a black-box manner so that to approximate the posterior distribution of interest as well as the marginal likelihood. The software is developed in \Cpp\ with interfaces with the softwares \R, \Matlab\ and \Octave.
}
\Plainkeywords{probabilistic programming, sequential Monte Carlo, particle filters, Markov chain Monte Carlo, particle MCMC, graphical models, BUGS, C++, R, Matlab, Octave} 

\Address{
  Adrien Todeschini\\
  INRIA Bordeaux Sud-Ouest\\
  200 avenue de la vieille tour\\
  33405 Talence Cedex, France
  E-mail: \email{Adrien.Todeschini@inria.fr}\\
  URL: \url{https://sites.google.com/site/adrientodeschini/}\medskip

  Fran\c cois Caron\\
  Department of Statistics\\
  University of Oxford\\
  1 South Parks Road\\
  OX13TG, Oxford, United Kingdom\\
  E-Mail: \email{caron@stats.ox.ac.uk}\\
  URL: \url{http://www.stats.ox.ac.uk/~caron/}
}

\begin{document}



\section{Introduction}
Bayesian inference aims at approximating the conditional probability law of an unknown parameter $X$ given some observations $Y$. Several problems such as signal filtering, object tracking or clustering can be cast into this framework. This conditional probability law is in general not analytically tractable. Markov chain Monte Carlo (MCMC) methods~\citep{Gilks1995,Robert2004}, and in particular Gibbs samplers, have been extensively used over the past 20 years in order to provide samples asymptotically distributed from the conditional distribution of interest.
As stated by~\cite{Cappe2000}
\begin{quote}
\textit{``The main factor in the success of MCMC algorithms is that they can be implemented with little effort in a large variety of settings. This is obviously true of the Gibbs sampler, which, provided some conditional distributions are available, simply runs by generating from these conditions, as shown by the BUGS software.''}
\end{quote}

The \BUGS\ (which stands for Bayesian Inference Using Gibbs Sampling) software has actually greatly contributed to the development of Bayesian and MCMC techniques among applied fields~\citep{Lunn2012}. \BUGS\ allows the user to define statistical models in a natural language, the \BUGS\ language~\citep{Gilks1994}, then approximates the posterior distribution of the parameter $X$ given the data using MCMC methods and provides some summary statistics. It is easy to use even for people not aware of MCMC methods and works as a black box. Various softwares have been developed based on or inspired by the `classic' \BUGS\ software, such as \WinBUGS, \OpenBUGS, \JAGS\ or \Stan.

A new generation of algorithms, based on {\bf interacting particle systems}, has generated a growing interest over the last 20 years. Those methods are known under the names of \textit{interacting MCMC, particle filtering, sequential Monte Carlo methods} (SMC)\footnote{Because of its widespread use in the Bayesian community, we will use the latter term in the remaining of this article.}. For some problems, those methods have shown to be more appropriate than MCMC methods, in particular for time series or highly correlated variables~\citep{Doucet2000,Doucet2001,Liu2001,DelMoral2004,Douc2014}. Contrary to MCMC methods, SMC do not require the convergence of the algorithm to some equilibrium and are particularly suited to dynamical estimation problems such as signal filtering or object tracking. Moreover, they provide unbiased estimates of the marginal likelihood at no additional computational cost. They have found numerous applications in signal filtering and robotics~\citep{Thrun2001,Gustafsson2002,Vermaak2002,Vo2003,Ristic2004,Caron2007}, systems biology~\citep{Golightly2006,Golightly2011,Bouchard-Cote2012}, economics and macro-economics~\citep{Pitt1999,Fernandez-Villaverde2007,Flury2011,DelMoral2012}, epidemiology~\citep{Cauchemez2008,Dureau2013}, ecology~\citep{Buckland2007,Peters2010} or pharmacology~\citep{Donnet2011}.  The introduction of the monograph of \cite{DelMoral2013} provides early references on this class of algorithms and an extensive list of application domains.

Traditionally, SMC methods have been restricted to the class of state-space models or hidden Markov chain models for which those models are particularly suited~\citep{Cappe2005}. However, SMC are far from been restricted to this class of models and have been used more broadly, either alone or as part of a MCMC algorithm \citep{Fearnhead2004,Fearnhead2007,Caron2008,Andrieu2010,Caron2012,Naesseth2014}.\medskip

The \Biips\ software\footnote{\href{http://alea.bordeaux.inria.fr/biips}{http://alea.bordeaux.inria.fr/biips}}, which stands for \emph{Bayesian Inference with Interacting Particle Systems}, has the following features:
\begin{itemize}
\item \textbf{\BUGS\ compatibility}: Similarly to the softwares \WinBUGS, \OpenBUGS\ and \JAGS, it allows users to describe complex statistical models in the \BUGS\ probabilistic language.
\item \textbf{Extensibility:} \R/\Matlab\ custom functions or samplers can be added to the \BUGS\ library.
\item \textbf{Black-box SMC inference:} It runs sequential Monte Carlo based algorithms (forward SMC, forward-backward SMC, particle independent Metropolis-Hastings, particle marginal Metropolis-Hastings) to provide approximations of the posterior distribution and of the  marginal likelihood.
\item \textbf{Post-processing:} The software provides some tools for extracting summary statistics (mean, variance, quantiles, etc.) on the variables of interest from the output of the SMC-based algorithms.
\item \textbf{\R/\Matlab/\Octave\ interfaces:} The software is developed in \Cpp\ with interfaces with the softwares \R, \Matlab\ and \Octave.
\end{itemize}

This article is organized as follows. Section~\ref{sec:graphical} describes the representation of the statistical model as a graphical model and the \BUGS\ language. Section~\ref{sec:SMC} provides the basics of SMC and particle MCMC algorithms. The main features of the \Biips\ software and its interfaces to \R\ and \Matlab/\Octave\ are given in Section~\ref{sec:biips}. Sections~\ref{sec:volatility} and~\ref{sec:kinetic} provide illustrations of the use of the software for Bayesian inference in stochastic volatility and stochastic kinetic models. In Section~\ref{sec:discussion} we discuss the relative merits and limits of \Biips\ compared to alternatives, and we conclude in Section~\ref{sec:conclusion}.

\section[Graphical models and BUGS language]{Graphical models and \BUGS\ language}
\label{sec:graphical}

\subsection{Graphical models}
A Bayesian statistical model is represented by a joint distribution $\mathcal{L}(X,Y)$ over the parameters $X$ and the observations $Y$. The joint distribution decomposes as $\mathcal{L}(X,Y)=\mathcal{L}(Y|X)\mathcal{L}(X)$ where the two terms of the right-hand side are respectively named \textit{likelihood} and \textit{prior}. As stated in the introduction, the objective of Bayesian inference is to approximate the posterior distribution $\mathcal{L}(X|Y=y)$ after having observed some data $y$.

A convenient way of representing a statistical model is through a directed acyclic graph~\citep{Lauritzen1996,Green2003,Jordan2004}. Such a graph provides at a glance the conditional independencies between variables and displays the decomposition of the joint distribution. As an example, consider the following switching stochastic volatility model \eqref{eq:volatilitymodel}.

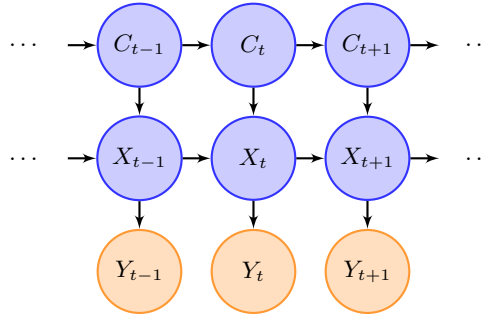
\begin{figure}[h]
\begin{center}
\begin{tikzpicture}[node distance=1.5cm,auto,>=latex',minimum size=1.1cm]
\node (c_t-2)   [] {\ldots};
\node (c_t-1)   [state,right of=c_t-2] {$C_{t-1}$};
\node (c_t)   [state,right of=c_t-1] {$C_t$};
\node (c_t+1)   [state,right of=c_t] {$C_{t+1}$};
\node (c_t+2)   [right of=c_t+1] {\ldots};
\node (X_t-2)   [below of=c_t-2] {\ldots};
\node (X_t-1)   [state,right of=X_t-2] {$X_{t-1}$};
\node (X_t)   [state,right of=X_t-1] {$X_t$};
\node (X_t+1)   [state,right of=X_t] {$X_{t+1}$};
\node (X_t+2)   [right of=X_t+1] {\ldots};
 \node (Y_t)   [measurement,below of=X_t] {$Y_t$};
 \node (Y_t-1)   [measurement,below of=X_t-1] {$Y_{t-1}$};
 \node (Y_t+1)   [measurement,below of=X_t+1] {$Y_{t+1}$};
  \path[->] (c_t-1) edge[thick] (c_t);
 \path[->] (c_t-2) edge[thick] (c_t-1);
 \path[->] (c_t) edge[thick] (c_t+1);
 \path[->] (c_t+1) edge[thick] (c_t+2);
 \path[->] (X_t-1) edge[thick] (X_t);
 \path[->] (X_t-2) edge[thick] (X_t-1);
 \path[->] (X_t) edge[thick] (X_t+1);
 \path[->] (X_t+1) edge[thick] (X_t+2);
  \path[->] (c_t) edge[thick] (X_t);
 \path[->] (c_t-1) edge[thick] (X_t-1);
 \path[->] (c_t+1) edge[thick] (X_t+1);
 \path[->] (X_t) edge[thick] (Y_t);
 \path[->] (X_t-1) edge[thick] (Y_t-1);
 \path[->] (X_t+1) edge[thick] (Y_t+1);
 \end{tikzpicture}
 \caption{Graphical representation of the switching volatility model as a directed acyclic graph. An arrow from node $A$ to node $B$ indicates that $A$ is a parent of $B$. The set of parents of a given node $A$ is noted $\text{pa}(A)$. For example, $\text{pa}(X_t)=(C_t,X_{t-1})$. Blue nodes correspond to unobserved variables, orange nodes to observed variables.}
 \label{fig:graph}
 \end{center}
 \end{figure}

\begin{example2}[Switching stochastic volatility]
 Let $Y_t$ be the response variable (log-return) and $X_t$ the unobserved
 log-volatility of $Y_t$. The stochastic volatility model is defined as follows
 for $t=1,\ldots, t_{\max}$
 \begin{subequations}
 \label{eq:volatilitymodel}
\begin{align}
X_t|(X_{t-1}=x_{t-1},C_t=c_t) &\sim \mathcal N (\alpha_{c_t} + \phi x_{t-1} , \sigma^2)\\
Y_t|X_t=x_t &\sim \mathcal N (0, \exp(x_t))
 \end{align}
where  `$\sim$' means `statistically distributed from', $\mathcal N(m,\sigma^2)$ denotes the normal
 distribution of mean $m\in \mathbb R$ and variance $\sigma^2>0$, $\alpha_1,\alpha_2\in\mathbb R$ and $\phi\in[-1,1]$. The regime variables $C_t\in\{1,2\}$ follow a two-state Markov process with
 transition probabilities
\begin{align}
 p_{ij}=\Pr(C_t=j|C_{t-1}=i)
 \end{align}
 for $i,j=1,2$ with $0<p_{ij}<1$ and $p_{i1}+p_{i2}=1$.
 \end{subequations}
 The graphical representation of the switching volatility model as a directed acyclic graph is given in Figure~\ref{fig:graph}.
\end{example2}

\subsection[BUGS language]{\BUGS\ language}

The \BUGS\ language is a probabilistic programming language that allows to define a complex stochastic model by decomposing the model into simpler conditional distributions~\citep{Gilks1994}. We refer the reader to the \JAGS\ user manual~\citep{Plummer2012} for details on the \BUGS\ language. The transcription of the switching stochastic volatility model~\eqref{eq:volatilitymodel} in \BUGS\ language is given in Listing~\ref{listing:sswbugs}\footnote{We truncated the Gaussian transition on $x_t$ to lie in the interval $[-500,500]$ in order to prevent the measurement precision $\exp(-x_t)$ to be numerically approximated to zero, which would produce an error.}.

\begin{lstlisting}[style=bugs,caption=Switching stochastic volatility model in BUGS language,label=listing:sswbugs,float]
model
{
  c[1] ~ dcat(pi[c0,])
  mu[1] <- alpha[1] * (c[1] == 1) + alpha[2] * (c[1] == 2) + phi * x0
  x[1] ~ dnorm(mu[1], 1/sigma^2) T(-500,500)
  y[1] ~ dnorm(0, exp(-x[1]))
  for (t in 2:t_max)
  {
    c[t] ~ dcat(ifelse(c[t-1] == 1, pi[1,], pi[2,]))
    mu[t] <- alpha[1] * (c[t] == 1) + alpha[2] * (c[t] == 2) + phi * x[t-1]
    x[t] ~ dnorm(mu[t], 1/sigma^2) T(-500,500)
    y[t] ~ dnorm(0, exp(-x[t]))
  }
}
\end{lstlisting}

\section{Sequential Monte Carlo methods}
\label{sec:SMC}
\subsection{Ordering and arrangement of the nodes in the graphical model}
\label{sec:ordering}
\begin{figure}
\begin{center}
\subfigure[Graphical model before rearrangement]{
\begin{tikzpicture}[node distance=1.3cm,auto,>=latex',scale=.2]
\node (X_1)   [state] {$X_1$};
\node (Y_1)   [measurement,right of=X_1] {$Y_{1}$};
\node (X_2)   [state,right of=Y_1] {$X_2$};
\node (Y_4)   [measurement,right of=X_2] {$Y_4$};
\node (Y_3)   [measurement,below of=X_1] {$Y_3$};
 \node (X_3)   [state,above of=X_2] {$X_3$};
 \node (Y_4)   [measurement,right of=X_2] {$Y_{4}$};
 \node (Y_2)   [measurement,below of=X_2] {$Y_2$};
 \path[->] (X_1) edge[thick] (Y_3);
 \path[->] (X_1) edge[thick,style={bend left}] (X_3);
 \path[->] (X_1) edge[thick] (Y_1);
 \path[->] (Y_1) edge[thick] (X_2);
 \path[->] (X_3) edge[thick] (X_2);
 \path[->] (X_2) edge[thick] (Y_2);
 \path[->] (X_2) edge[thick] (Y_4);
 \end{tikzpicture}}\hskip 1cm
 \subfigure[Topological sort (with priority to measurement nodes): $(X_1,Y_1,Y_3,X_3,X_2,Y_4,Y_2)$. Note that this sort is not unique.]{
\begin{tikzpicture}[node distance=1.3cm,auto,>=latex',scale=.2]
\node (X_1)   [state] {1};
\node (Y_1)   [measurement,right of=X_1] {2};
\node (X_2)   [state,right of=Y_1] {5};
\node (Y_3)   [measurement,below of=X_1] {3};
 \node (X_3)   [state,above of=X_2] {4};
 \node (Y_4)   [measurement,right of=X_2] {6};
 \node (Y_2)   [measurement,below of=X_2] {7};
 \path[->] (X_1) edge[thick] (Y_3);
 \path[->] (X_1) edge[thick,style={bend left}] (X_3);
 \path[->] (X_1) edge[thick] (Y_1);
 \path[->] (Y_1) edge[thick] (X_2);
 \path[->] (X_3) edge[thick] (X_2);
 \path[->] (X_2) edge[thick] (Y_2);
 \path[->] (X_2) edge[thick] (Y_4);
 \end{tikzpicture}}\hskip 1cm
\subfigure[Graphical model after rearrangement]{
\begin{tikzpicture}[node distance=1.3cm,auto,>=latex']
\node (X_0)   [state] {$X_1'$};
\node (Y_0)   [measurement,right of=X_0] {$Y_{1}'$};
\node (X_1)   [state,right of=Y_0] {$X_2'$};
\node (Y_1)   [measurement,below of=X_1] {$Y_2'$};
 \path[->] (X_0) edge[thick] (Y_0);
 \path[->] (X_0) edge[thick,style={bend left=60}] (X_1);
 \path[->] (Y_0) edge[thick] (X_1);
 \path[->] (X_1) edge[thick] (Y_1);
 \end{tikzpicture}}
 \caption{Rearrangement of a directed acyclic graph. $X_1'=X_1$, $Y_1'=\{Y_1,Y_3\}$, $X_2'=(X_3,X_2)$ and $Y_2'=\{Y_2,Y_4\}$. The statistical model decomposes as $p(x_1',x_2',y_1',y_2')=p(x_1')p(y'_1|x'_1)p(x_2'|x_1',y_1')p(y_2'|x_2')$.}
 \label{fig:graphorder}
 \end{center}
\end{figure}
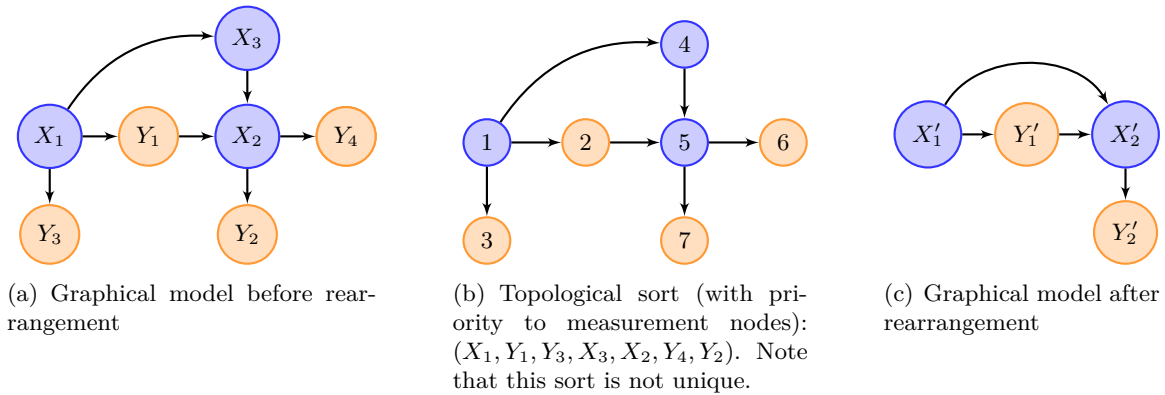

In order to apply sequential Monte Carlo methods in an efficient manner, \Biips\ proceeds to a rearrangement of the nodes of the graphical model as follows:
\begin{enumerate}
\item Sort the nodes of the graphical model according to a topological order (parents nodes before children), by giving priority to measurement nodes compared to state nodes (note that the sort is not unique);
\item Group together successive measurement or state nodes;
\item We then obtain an ordering $(X_1,Y_1,X_2,Y_2,\ldots,X_n,Y_n)$ where $X_i$ correspond to groups of unknown variables, and $Y_i$ to groups of observations.
\end{enumerate}
Figure \ref{fig:graphorder} gives an example of rearrangement of a graphical model.

\begin{example2}[Switching stochastic volatility model (continued)]
For the model \eqref{eq:volatilitymodel}, the graphical model can be decomposed as $X_t'=(X_t,C_t)$, $Y_t'=Y_t$ and $n=t_{\max}$. In this particular case, the resulting graphical model is a hidden Markov model. Note however that it does not have to be the case in general, as illustrated in Figure~\ref{fig:graphorder}.
\end{example2}

\subsection{Sequential Monte Carlo algorithm}
Assume that we have variables $(X_1,Y_1,\ldots,X_n,Y_n)$ which are sorted as described in Section~\ref{sec:ordering}, where $X_i$ and $Y_i$ respectively correspond to unobserved and observed variables, for $i=1,\ldots,n$. By convention, let $X_{a:b}=(X_a,X_{a+1},\ldots,X_b)$, $a< b$. Also, let $X_{t}\in\mathcal X'_t$ and $X_{1:t}\in\mathcal X_t$ for $t=1,\ldots,n$ where $\mathcal X_t= \mathcal X_{t-1} \otimes \mathcal X'_t$. The statistical model decomposes as
 \begin{equation}
p(x_{1:n},y_{1:n})=p(x_1)p(y_1|x_1)\prod_{t=2}^n p(x_t|\text{pa}(x_t))p(y_t|\text{pa}(y_t))
\end{equation}
where pa$(x)$ denotes the set of parents of variable $x$ in the decomposition described in Section~\ref{sec:ordering}.

Sequential Monte Carlo methods~\citep{Doucet2001,DelMoral2004,Doucet2011} proceed by sequentially approximating conditional distributions
\begin{equation}
\pi_t(x_{1:t}|y_{1:t})= \frac{p(x_{1:t},y_{1:t})}{p(y_{1:t})}
\end{equation}

for $t=1,\ldots,n$, by a weighted set of $N$ particles $(X_{t,1:t}^{(i)},W_t^{(i)})_{i=1,\ldots,N}$ that evolve according to two mechanisms:
\begin{itemize}
\item \textbf{Mutation/Exploration}: Each particle $i$ is randomly extended with $X_{t,t+1}^{(i)}$
\item \textbf{Selection}: Each particle is associated a weight $W_t^{(i)}$ depending on its fit to the data. Particles with high weights are duplicated while particles with low weights are deleted.
\end{itemize}

The vanilla sequential Monte Carlo algorithm is given in Algorithm \ref{algo:smc}.

\begin{algorithm}
\caption{Standard sequential Monte Carlo algorithm}
\label{algo:smc}
$\bullet$ For $t=1,\ldots,n$

$\qquad \bullet$ For $i=1,\ldots,N$, sample $X_{t,t}^{(i)}\sim q_t$ and let $X_{t,1:t}^{(i)} =(\widetilde X_{t-1,1:t-1}^{(i)},X_{t,t}^{(i)}) $

$\qquad \bullet$ For $i=1,\ldots,N$, set
$$
w_t^{(i)}=\frac{\pi(y_t|\text{pa}(y_t))\pi({x}_{t,t}^{(i)}|\text{pa}({x}_{t,t}^{(i)}))}{q_t({x}_{t,t}^{(i)})}
$$

$\qquad \bullet$ For $i=1,\ldots,N$, set $W_t^{(i)}=\frac{w_t^{(i)}}{\sum_{j=1}^Nw_t^{(j)}}$

$\qquad \bullet$ Duplicate particles of high weight and delete particles of low weight using some resampling strategy. Let $\widetilde X_{t,1:t}^{(i)}$, $i=1,\ldots,N$ be the resulting set of particles with weights $\frac{1}{N}$.

$\bullet$ Outputs:

$\qquad \bullet$ Weighted particles $(W_t^{(i)},X_{t,1:t}^{(i)})_{i=1,\ldots,N}$ for $t=1,\ldots,n$

$\qquad \bullet$ Estimate of the marginal likelihood $\widehat Z = \prod_{t=1}^n \left (\frac{1}{N}\sum_{i=1}^N w_t^{(i)} \right )$
\end{algorithm}

The output of the algorithm is a sequence of weighted particles providing approximations of the successive conditional distributions $\pi_t$. In particular, $\left (W_n^{(i)},X_{n,1:n}^{(i)}\right )_{i=1,\ldots,N}$ provides a particle approximation of the full conditional distribution $\pi_n(x_{1:n}|y_{1:n})$ of the unknown variables given the observations. Point estimates of the parameters can then be obtained. For any function $h:\mathcal X_n\rightarrow S$
\begin{align}
\mathbb E[h(X_{1:n})|Y_{1:n}=y_{1:n}]\simeq \sum_{i=1}^N W_{n}^{(i)} h(X_{n,1:n}^{(i)})
\end{align}
For example, by taking $h(X_{1:n})=X_{1:n}$ one obtains posterior mean estimates
\begin{align*}
\mathbb E[X_{1:n}|Y_{1:n}=y_{1:n}]\simeq \sum_{i=1}^N W_{n}^{(i)} X_{n,1:n}^{(i)}.
\end{align*}

The algorithm also provides an unbiased estimate of the marginal likelihood
\begin{align}
p(y_{1:n})=\int_{\mathcal X_n} p(y_{1:n},x_{1:n})dx_{1:n}\simeq \widehat Z
\end{align}

$q_t$ is the proposal/importance density function and is used for exploration. This proposal may be a function of \text{pa}$(x_t)$ and/or $y_t$. The simplest is to use the conditional distribution $\pi(x_t|\text{pa}(x_t))$, which is directly given by the statistical model, as a proposal distribution. A better choice is to use the distribution $\pi(x_t|\text{pa}(x_t), y_t)$, or any approximation of this distribution.

\subsection{Limitations of SMC algorithms and diagnostic}

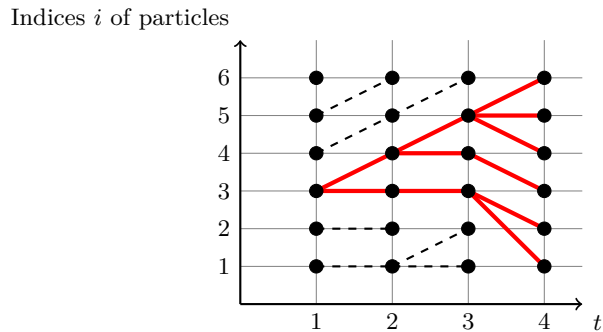
\begin{figure}
\begin{center}
\begin{tikzpicture}
\draw[gray,very thin]  (1,0) node[black,anchor=north] {1} -- (1,3.5);
\draw[gray,very thin]  (2,0) node[black,anchor=north] {2} -- (2,3.5);
\draw[gray,very thin]  (3,0) node[black,anchor=north] {3} -- (3,3.5);
\draw[gray,very thin]  (4,0) node[black,anchor=north] {4} -- (4,3.5);

\draw[gray,very thin]  (0,.5) node[black,anchor=east] {1} -- (4.5,.5);
\draw[gray,very thin]  (0,1) node[black,anchor=east] {2} -- (4.5,1);
\draw[gray,very thin]  (0,1.5) node[black,anchor=east] {3} -- (4.5,1.5);
\draw[gray,very thin]  (0,2) node[black,anchor=east] {4} -- (4.5,2);
\draw[gray,very thin]  (0,2.5) node[black,anchor=east] {5} -- (4.5,2.5);
\draw[gray,very thin]  (0,3) node[black,anchor=east] {6} -- (4.5,3);

\draw[thick,->] (0,0) -- (4.5,0) node[anchor=north west] {$t$};
\draw[thick,->] (0,0) -- (0,3.5) node[anchor=south east] {Indices $i$ of particles};
\draw[thick,dashed] (1, .5) -- (2, .5);
\draw[thick,dashed] (1, 1) -- (2, 1);
\draw[ultra thick,red] (1, 1.5) -- (2, 1.5);
\draw[ultra thick,red] (1, 1.5) -- (2, 2);
\draw[thick,dashed] (1, 2) -- (2, 2.5);
\draw[thick,dashed] (1, 2.5) -- (2, 3);
\draw[thick,dashed] (2, .5) -- (3, .5);
\draw[thick,dashed] (2, .5) -- (3, 1);
\draw[ultra thick,red] (2, 1.5) -- (3, 1.5) ;
\draw[ultra thick,red] (2, 2) -- (3, 2);
\draw[ultra thick,red] (2, 2) -- (3, 2.5);
\draw[thick,dashed] (2, 2.5) -- (3, 3);
\draw[ultra thick,red] (3, 1.5) -- (4, 0.5);
\draw[ultra thick,red] (3, 1.5) -- (4, 1);
\draw[ultra thick,red] (3, 2) -- (4, 1.5);
\draw[ultra thick,red] (3, 2.5) -- (4, 2);
\draw[ultra thick,red] (3, 2.5) -- (4, 2.5);
\draw[ultra thick,red] (3, 2.5) -- (4, 3);

\draw[black,fill] (1,1) circle (2.5pt);
\draw[black,fill] (1,1.5) circle (2.5pt);
\draw[black,fill] (1,2) circle (2.5pt);
\draw[black,fill] (1,2.5) circle (2.5pt);
\draw[black,fill] (1,3) circle (2.5pt);
\draw[black,fill] (1,0.5) circle (2.5pt);

\draw[black,fill] (2,1) circle (2.5pt);
\draw[black,fill] (2,1.5) circle (2.5pt);
\draw[black,fill] (2,2) circle (2.5pt);
\draw[black,fill] (2,2.5) circle (2.5pt);
\draw[black,fill] (2,3) circle (2.5pt);
\draw[black,fill] (2,0.5) circle (2.5pt);

\draw[black,fill] (3,1) circle (2.5pt);
\draw[black,fill] (3,1.5) circle (2.5pt);
\draw[black,fill] (3,2) circle (2.5pt);
\draw[black,fill] (3,2.5) circle (2.5pt);
\draw[black,fill] (3,3) circle (2.5pt);
\draw[black,fill] (3,0.5) circle (2.5pt);

\draw[black,fill] (4,1) circle (2.5pt);
\draw[black,fill] (4,1.5) circle (2.5pt);
\draw[black,fill] (4,2) circle (2.5pt);
\draw[black,fill] (4,2.5) circle (2.5pt);
\draw[black,fill] (4,3) circle (2.5pt);
\draw[black,fill] (4,0.5) circle (2.5pt);
\end{tikzpicture}
\end{center}
\caption{Genealogical tree of a sequential Monte Carlo algorithm. A line from an index $i$ at time $t$ to an index $j$ at time $t+1$ indicates that $j$ is a children of $i$. Dashed lines correspond to particles which were deleted at some stage $t<n$.  For example, child$(i=5,t=3)=\{4,5,6\}$, child$(i=6,t=2)=\emptyset$, anc$(i=3,t=4,1)=4$, a(4,1)=\{3\}, a(4,2)=\{3,4\}. The SESS in this particular example takes the following values: $\SESS(4,4)=\left [ \left ( W_{4}^{(1)}\right )^2 + \left ( W_{4}^{(2)}\right )^2 +\left ( W_{4}^{(3)}\right )^2+\left ( W_{4}^{(4)}\right )^2+\left ( W_{4}^{(5)}\right )^2+\left ( W_{4}^{(6)}\right )^2\right ]^{-1}$, $\SESS(4,3)=\left [\left ( W_{4}^{(1)}+W_{4}^{(2)}\right )^2 +
 \left ( W_{4}^{(3)}\right )^2+ \left ( W_{4}^{(4)}+W_{4}^{(5)}+W_{4}^{(6)}\right )^2  \right ]^{-1}$, $\SESS(4,2)=\left [ \left ( W_{4}^{(1)}+W_{4}^{(2)}\right )^2 + \left ( W_{4}^{(3)}+W_{4}^{(4)}+W_{4}^{(5)} + W_{4}^{(6)}\right )^2  \right ]^{-1}$ and $\SESS(4,1)=1$.}
\label{fig:SESS}
\end{figure}

Due to the successive resampling, the quality of the particle approximation of $p(x_{t:n}|y_{1:n})$ decreases as $n-t$ increases, a problem referred as sample degeneracy or impoverishment, see e.g.~\cite{Doucet2011}. \Biips\ uses a simple criteria to provide a diagnostic on the output of the SMC algorithm.

Let child$(i,t)\subseteq \{1,\ldots N\}$ be set of indices of the children of particle $i$ at time $t$. Note that if a particle $i$ is deleted at time $t$, then child$(i,t)=\emptyset$. Similarly, let anc$(i,t,1)\in \{1,\ldots N\}$ be the index of the first-generation ancestor of particle $i$ at time $t$. We therefore have  $i\in \text{child}(\text{anc}(i,t+1,1), t)$.  By extension, we write anc$(i,t,2)$ for the index of the second-generation ancestor of particle $i$ at times $t$. Let $a(n,t)\subseteq \{1,\ldots N\}$ be the set of unique values in (anc$(i,n,n-t))_{i=1,\ldots,N}$. Due to the successive resampling, the number of unique ancestors $K_{n,t}=\text{card}(a(n,t))$ of the particles at time $n$ decreases as $n-t$ increases. A measure of the quality of the approximation of the marginal posterior distributions $p(x_{t:n}|y_{1:n})$, for $1\leq t\leq n$, is given by the smoothing effective sample size ($\SESS$):
\begin{equation}
\SESS(n,t)=\left [ \sum_{j\in a(n,t)}\left (\sum_{i| \text{anc}(i,n,n-t)=j} W_{n}^{(i)}\right )^2 \right ]^{-1}
\end{equation}
with $1\leq\SESS(n,t) \leq N$. Figure~\ref{fig:SESS} provides an illustration on a simple example. Larger values of the SESS indicate better approximation. As explained earlier, this value is likely to decrease with the number $n-t$ due to the successive resamplings. For a given value of $n$, one can increase the number of particles $N$ in order to obtain an acceptable SESS. In a simple importance sampling framework, the ESS corresponds to the number of perfect samples from the posterior needed to obtain an estimator with similar variance~\citep{Doucet2011}; as a rule of thumb, the minimal value is set to $30$.

Nonetheless, in cases where $n$ is very large, or when a given unobserved node is the parent of a large number of other unobserved nodes (\textit{sometimes referred as parameter estimation problem in sequential Monte Carlo}), this degeneracy may be too severe in order to achieve acceptable results. To address such limitations, \cite{Andrieu2010} have recently proposed a set of techniques for mitigating SMC algorithms with MCMC methods by using the former as a proposal distribution. We present such algorithms in the next section.

\subsection{Particle MCMC}
\label{sec:pmcmc}
One of the pitfalls of sequential Monte Carlo is that they suffer from degeneracy due to the successive resamplings. For large graphical models, or for graphical models where some variables have many children nodes, the particle approximation of the full posterior will be poor. Recently, algorithms have been developed that propose to use SMC algorithms within a MCMC algorithm~\citep{Andrieu2010}. The particle independent Metropolis-Hastings (PIMH) algorithm~\ref{algo:pimh} provides MCMC samples $(X_{1:n}(k))_{k=1,\ldots,n_\text{iter}}$ asymptotically distributed from the posterior distribution, using a SMC algorithm as proposal distribution in an independent Metropolis-Hastings (MH) algorithm.

\begin{algorithm}
\caption{Standard particle independent Metropolis-Hastings algorithm}
\label{algo:pimh}
Set $\widehat Z(0)=0$

$\bullet$ For $k=1,2,\ldots,n_\text{iter}$

$\qquad \bullet$ Run a sequential Monte Carlo algorithm to approximate $\pi_n(x_{1:n}|y_{1:n})$. \\
$\qquad\qquad$ Let $(X_{1:n}^{\star(i)},W_{n}^{\star(i)})_{i=1,\ldots,N}$ and $\widehat Z^\star$ be  respectively the set of weighted particles and the estimate of the marginal likelihood.

$\qquad \bullet$ With probability
\begin{equation*}
\min\left (1,\frac{\widehat Z^\star}{\widehat Z(k-1)} \right )
\end{equation*}
$\qquad\qquad \bullet$ set $X_{1:n}(k)=X_{1:n}^{\star(\ell)}$ and $\widehat Z(k-1)=\widehat Z^\star$,  where $\ell\sim\Discrete(W_{n}^{\star(1)},\ldots,W_{n}^{\star(N)})$

$\qquad\qquad \bullet$ otherwise, set $X_{1:n}(k)=X_{1:n}(k-1)$ and $\widehat Z(k)=\widehat Z(k-1)$

$\bullet$ Output:

$\qquad \bullet$ MCMC samples $(X_{1:n}(k))_{k=1,\ldots,n_\text{iter}}$
\end{algorithm}

The particle marginal Metropolis-Hastings (PMMH) algorithm splits the variables in the graphical model into two sets: one set of variables $X$ that will be sampled using a SMC algorithm, and a set $\theta=(\theta_1,\ldots,\theta_p)$ sampled with a MH proposal. It outputs MCMC samples $(X_{1:n}(k),\theta(k))_{k=1,\ldots,n_{\text{iter}}}$ asymptotically distributed from the posterior distribution. Algorithm~\ref{algo:pmmh} provides a description of the PMMH algorithm.

\begin{algorithm}
\caption{Standard particle marginal Metropolis-Hastings algorithm}
\label{algo:pmmh}
Set $\widehat Z(0)=0$ and initialize $\theta(0)$

$\bullet$ For $k=1,2,\ldots,n_\text{iter}$

$\qquad \bullet$ Sample $\theta^\star\sim \nu(\cdot|\theta(k-1))$

$\qquad \bullet$ Run a sequential Monte Carlo algorithm to approximate $\pi_n(x_{1:n}|y_{1:n},\theta^\star)$. Let $(X_{1:n}^{\star(i)},W_{n}^{\star(i)})_{i=1,\ldots,N}$ and $\widehat Z^\star$ be respectively the set of weighted particles and the estimate of the marginal likelihood.

$\qquad \bullet$ With probability
\begin{equation*}
\min\left (1,\frac{\widehat Z^\star }{\widehat Z(k-1)}\times \frac{p(\theta^\star)}{ p(\theta(k-1))}  \times \frac{\nu(\theta(k-1)|\theta^\star)}{\nu(\theta^\star|\theta(k-1))} \right )
\end{equation*}
$\qquad\qquad \bullet$ set $X_{1:n}(k)=X_{1:n}^{\star(\ell)}$, $\theta(k)=\theta^{\star}$ and $\widehat Z(k-1)=\widehat Z^\star$,  where $\ell\sim\Discrete(W_{n}^{\star(1)},\ldots,W_{n}^{\star(N)})$

$\qquad\qquad \bullet$ otherwise, set $\theta(k)=\theta(k-1)$, $X_{1:n}(k)=X_{1:n}(k-1)$ and $\widehat Z(k)=\widehat Z(k-1)$

$\bullet$ Output:

$\qquad \bullet$ MCMC samples $(X_{1:n}(k),\theta(k))_{k=1,\ldots,n_\text{iter}}$

\end{algorithm}

\section{Biips software}
\label{sec:biips}

\begin{figure}
\newcommand*{\vcenteredhbox}[1]{\begingroup
\setbox0=\hbox{#1}\parbox{\wd0}{\box0}\endgroup}

\begin{center}
\tikzstyle{script}=[rectangle, rounded corners,
                                    thick,
                                    text width=3cm,
                                    draw=blue!80,
                                    fill=blue!20,
                                    text centered,
                                    font=\large,
                                    anchor=mid, minimum height=2em]
\tikzstyle{cpp}=[rectangle, rounded corners,
                                    thick,
                                    text width=3cm,
                                    draw=orange!80,
                                    fill=orange!25,
                                    text centered,
                                    font=\large,
                                    anchor=mid, minimum height=2em]

\tikzstyle{inter}=[
                                    anchor=mid, 
                                    ]

\begin{tikzpicture}[node distance=1.2cm,auto,>=latex']
\node (core)   [cpp] {Core};
\node (base)   [cpp,above of= core] {Base};
\node (comp)   [cpp,above of=base] {Compiler};
\node (rbiips)   [script,above of=comp, xshift=-2cm] {\Rbiips};
\node (matbiips)   [script,above of=comp, xshift=2cm] {\Matbiips};
\node  [left of=rbiips,text width=3cm, text centered,node distance=3cm, text=blue, font=\normalsize] {\R+\pkg{Rcpp}};
\node  [right of=matbiips,text width=3cm, text centered,node distance=3.5cm, text=blue, font=\normalsize] {\Matlab+MEX};
\node (r) [inter,above of=rbiips] {\includegraphics[width=.8cm]{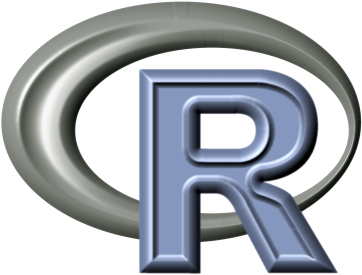}};
\node (matlab) [inter,above of=matbiips] {\includegraphics[width=.8cm]{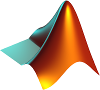}};
\node  [left of=base,node distance=3cm,text width=3cm, text centered,text=orange, text centered, text width=3cm, font=\normalsize] {\Cpp};
 \path[->] (base) edge[thick] (core);
 \path[->] (comp) edge[thick] (base);
 \path[->] (rbiips) edge[thick] (comp);
 \path[->] (matbiips) edge[thick] (comp);
 \path[->] (r) edge[thick] (rbiips);
 \path[->] (matlab) edge[thick] (matbiips);
\end{tikzpicture}
\end{center}
\caption{Biips architecture}
\label{fig:bricks}
\end{figure}
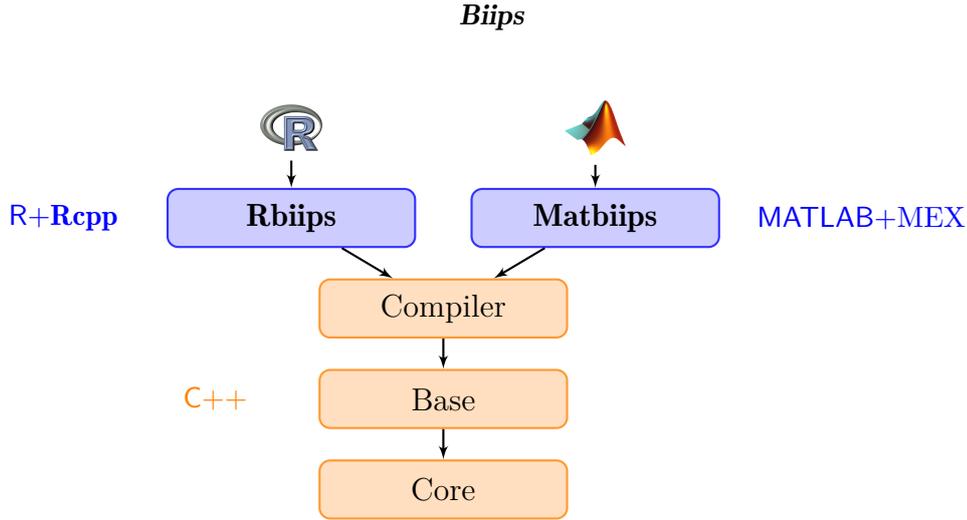

The \Biips\ code consists of three libraries written in \Cpp\, whose architecture is adapted from \JAGS\, and two user interfaces. Figure \ref{fig:bricks} summarizes the main components from the bottom to the top level.

\paragraph{\Biips\ \Cpp\ libraries}

\begin{itemize}
\item The \textit{Core} library contains the bottom level classes to represent a graphical model and run SMC algorithms.

\item The \textit{Base} library is an extensible collection of distributions, functions and samplers.

\item The \textit{Compiler} library allows to describe the model in \BUGS\ language and provides a controller for higher level interfaces.

\end{itemize}

\paragraph{\Biips\ user interfaces}

At the top level, we provide two user interfaces to the \Biips\ \Cpp\ classes:

\begin{itemize}
\item \Matbiips\ interface for \Matlab/\Octave.
\item \Rbiips\ interface for \R.
\end{itemize}

These interfaces for scientific programming languages make use of specific libraries that allow binding \Cpp\ code with their respective environment. \Matbiips\ uses the \Matlab\ MEX library or its \Octave\ analog. \Rbiips\ uses the \pkg{Rcpp} library~\citep{Eddelbuettel2011, Eddelbuettel2013}.

In addition, the interfaces provide user-friendly functions to facilitate the workflow for doing inference with \Biips. The typical workflow is the following:

\begin{enumerate}
\item Define the model and data.
\item Compile the model.
\item Run inference algorithms.
\item Diagnose and analyze the output.
\end{enumerate}
The main functions in \Matbiips\ and \Rbiips\ are described in Table~\ref{tab:functions}. Both interfaces use the same set of functions, with similar inputs/outputs; the interface is also similar to the \rjags~\citep{rjags} interface to \JAGS.
The \Matbiips\ interface does not use object-oriented programming due to compatibility issues with \Octave, whereas \Rbiips\ uses S3 classes. The prefix \code{biips_} is used for function names in order to avoid potential conflicts with other packages.

\begin{table}[h!]
\caption{List of the main functions in \Biips\ interfaces}
\label{tab:functions}
\begin{tabular}{ll}
\\
\hline
\multicolumn{2}{c}{\bf Construction of the model}\\
\hline
\code{biips_model} & Instantiates a BUGS-language stochastic model\\
\code{biips_add_function} & Adds a custom function\\
\code{biips_add_distribution} & Adds a custom sampler\medskip\\
\hline
\multicolumn{2}{c}{\bf Inference algorithms}\\
\hline
\code{biips_smc_samples} & Runs a SMC algorithm\\
\code{biips_smc_sensitivity} & Estimates the marginal likelihood for a set of parameter values\smallskip\\
\code{biips_pimh_init} & Initializes the PIMH\\
\code{biips_pimh_update} & Runs the PIMH (burn-in)\\
\code{biips_pimh_samples} & Runs the PIMH and returns samples\smallskip \\
\code{biips_pmmh_init} & Initializes the PMMH\\
\code{biips_pmmh_update} & Runs the PMMH (adaptation and burn-in)\\
\code{biips_pmmh_samples} & Runs the PMMH and returns samples\medskip\\
\hline
\multicolumn{2}{c}{\bf Diagnosis and summary}\\
\hline
\code{biips_diagnosis} & Performs a diagnosis of the SMC algorithm\\
\code{biips_density} & Returns kernel density estimates of the posterior (continuous)\\
\code{biips_table} & Returns probability mass estimates of the posterior (discrete)\\
\code{biips_summary} & Returns summary statistics of the posterior\\
\end{tabular}
\end{table}


\paragraph{Extensions of the \BUGS\ language with custom functions}

In addition to the user interfaces, we provide a simple way of extending the \BUGS\ language by adding custom distributions and functions. This is done by calling \Matlab/\Octave\ or \R\ functions from the \Cpp\ layer using the \proglang{C} \pkg{MEX} library in \Matlab/\Octave\ and the \pkg{Rcpp} package in \R.


\section{Example: Switching stochastic volatility model}
\label{sec:volatility}

\subsection{Bayesian inference with SMC}

\begin{figure}
\begin{center}
\includegraphics[width=.5\textwidth]{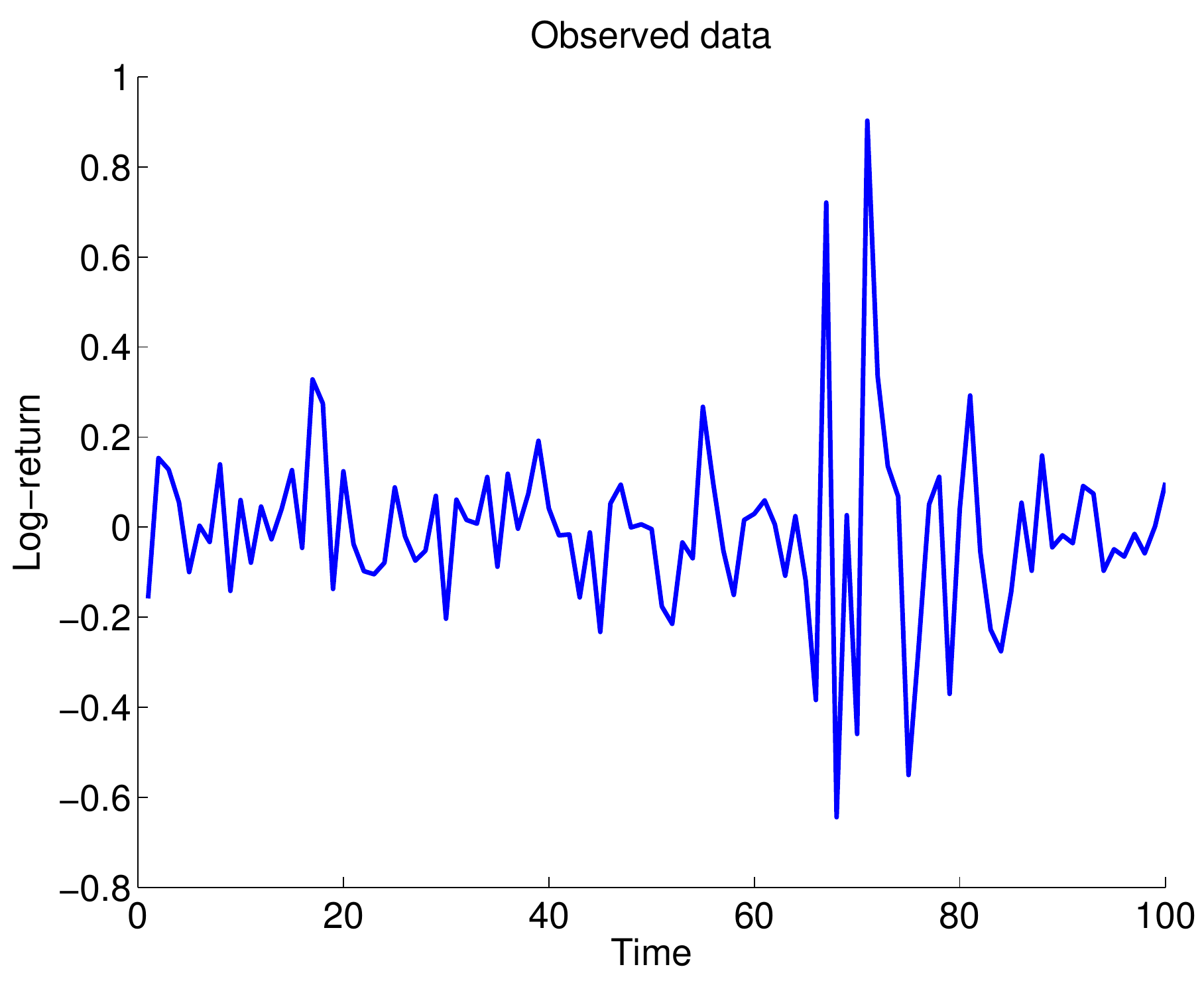}
\end{center}
\caption{Sampled data $y_1,\ldots,y_{t_{\max}}$ for the switching stochastic volatility model.}
\label{fig:volatility_obs}
\end{figure}

We consider the switching stochastic volatility model \eqref{eq:volatilitymodel}. Our objective is to approximate the filtering distributions $p(x_t|y_{1:t})$ and smoothing distributions $p(x_t|y_{1:t_{\max}})$, for $t=1,\ldots,t_{\max}$ and obtain some point estimates, such as the posterior means or posterior quantiles. The function \code{biips_model} parses and compiles the \BUGS\ model, and sample the data if \code{sample_data} is set to true. The data are represented in Figure~\ref{fig:volatility_obs}.

\begin{lstlisting}[style=matbiips]
sigma = .4; alpha = [-2.5; -1]; phi = .5; c0 = 1; x0 = 0; t_max = 100;
pi = [.9, .1; .1, .9];
data = struct('t_max', t_max, 'sigma', sigma,...
        'alpha', alpha, 'phi', phi, 'pi', pi, 'c0', c0, 'x0', x0);
model_file = 'switch_stoch_volatility.bug';
model = biips_model(model_file, data, 'sample_data', true);
data = model.data;
\end{lstlisting}

\begin{lstlisting}[style=rbiips]
sigma <- .4; alpha <- c(-2.5, -1); phi <- .5; c0 <- 1; x0 <- 0; t_max <- 100
pi <- matrix(c(.9, .1, .1, .9), nrow=2, byrow=TRUE)
data <- list(t_max=t_max, sigma=sigma,
             alpha=alpha, phi=phi, pi=pi, c0=c0, x0=x0)
model_file <- 'switch_stoch_volatility.bug'
model <- biips_model(model_file, data, sample_data=TRUE)
data <- model$data()
\end{lstlisting}

One can then run a sequential Monte Carlo algorithm with the function  \code{biips_smc_samples} to provide a particle approximation of the posterior distribution. The particle filter can be run in filtering, smoothing and/or backward smoothing modes, see~\citep{Doucet2011} for more details. By default, \Biips\ also automatically chooses the proposal distribution $q_t$.
\newpage

\begin{lstlisting}[style=matbiips]
n_part = 5000;
variables = {'x'};
out_smc = biips_smc_samples(model, variables, n_part);
diag_smc = biips_diagnosis(out_smc);
\end{lstlisting}

\begin{lstlisting}[style=rbiips]
n_part <- 5000
variables <- c('x')
out_smc <- biips_smc_samples(model, variables, n_part)
diag_smc <- biips_diagnosis(out_smc)
\end{lstlisting}

\begin{figure}
\begin{center}
\subfigure[Particles (smoothing)]{\includegraphics[width=.48\textwidth]{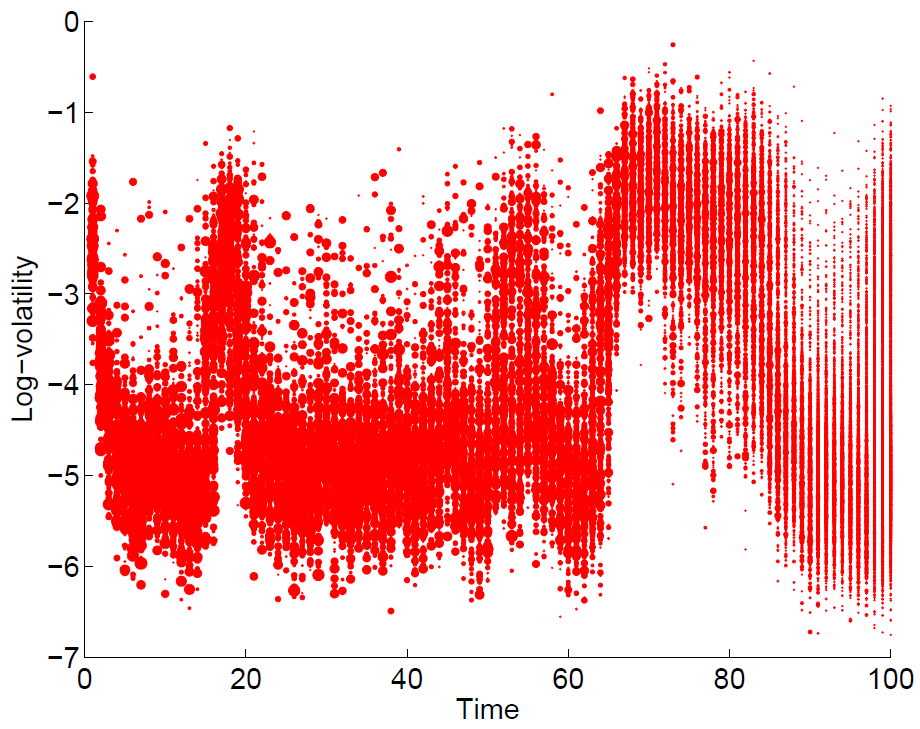}}
\subfigure[SESS]{\includegraphics[width=.48\textwidth]{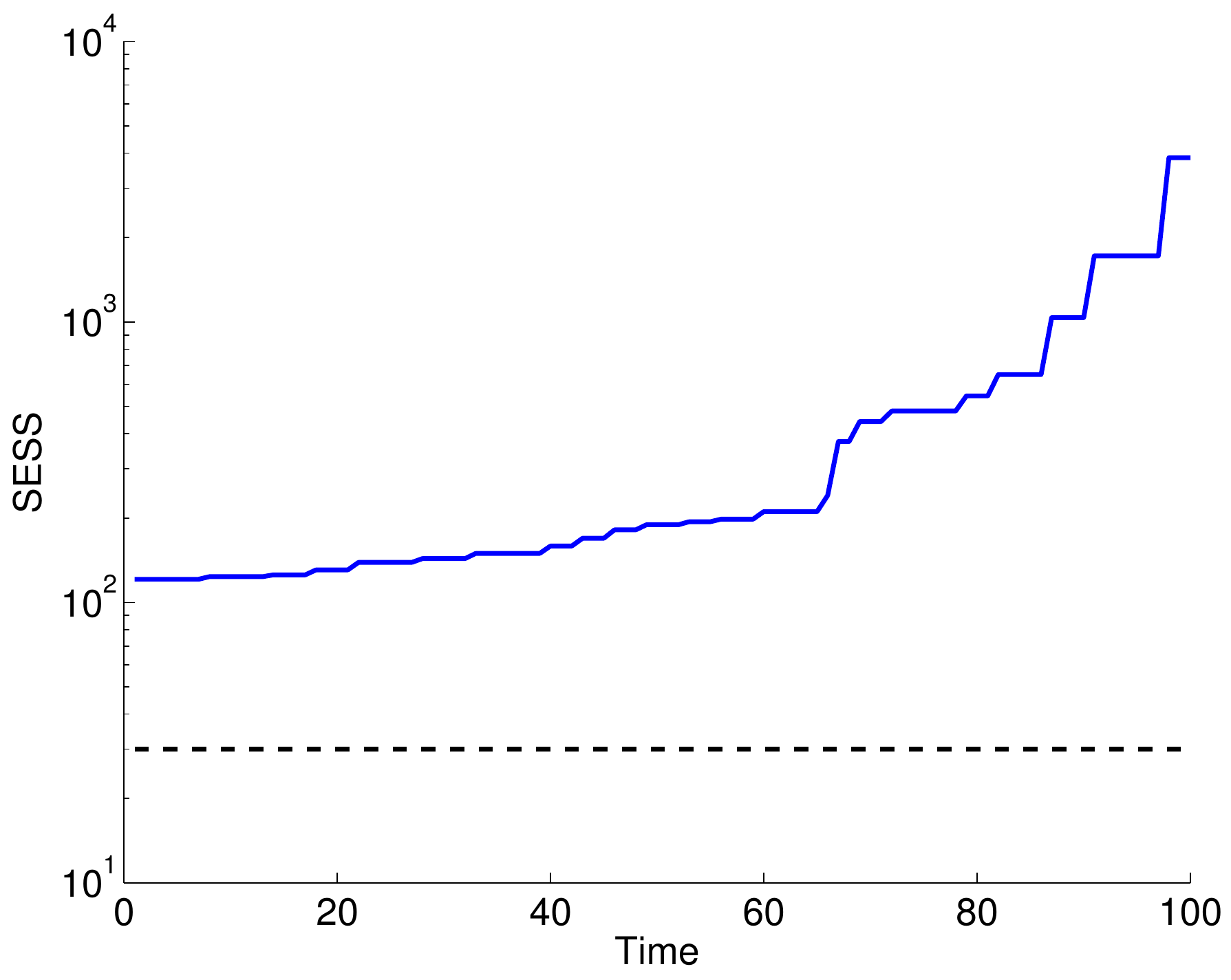}}
\end{center}
\caption{SMC: (a) Set of weighted particles of the posterior distribution for the switching stochastic volatility model. (b) Smoothing effective sample size with respect to $t$.}
\label{fig:volatility_particles}
\end{figure}

\code{out_smc} is an object containing the values of the particles and their weights for each of the monitored variables (the variable $X_{1:t_{\max}}$ in the example). An illustration of the weighted particles is given in Figure~\ref{fig:volatility_particles}(a). Figure~\ref{fig:volatility_particles}(b) shows the value of the SESS with respect to time. If the minimum is below the threshold of 30, \code{biips_diagnosis} recommends to increase the number of particles.


The function \code{biips_summary} provides some summary statistics on the marginal distributions (mean, quantiles, etc.), and \code{biips_density} returns kernel density estimates of the marginal posterior distributions.

\begin{lstlisting}[style=matbiips]
summ_smc = biips_summary(out_smc, 'probs', [.025, .975]);
x_f_mean = summ_smc.x.f.mean; x_f_quant = summ_smc.x.f.quant;
x_s_mean = summ_smc.x.s.mean; x_s_quant = summ_smc.x.s.quant;
kde_smc = biips_density(out_smc);
\end{lstlisting}

\begin{lstlisting}[style=rbiips]
summ_smc <- biips_summary(out_smc, probs=c(.025, .975))
x_f_mean <- summ_smc$x$f$mean; x_f_quant <- summ_smc$x$f$quant
x_s_mean <- summ_smc$x$s$mean; x_s_quant <- summ_smc$x$s$quant
kde_smc <- biips_density(out_smc)
\end{lstlisting}

\begin{figure}
\begin{center}
\subfigure[Filtering]{\includegraphics[width=.48\textwidth]{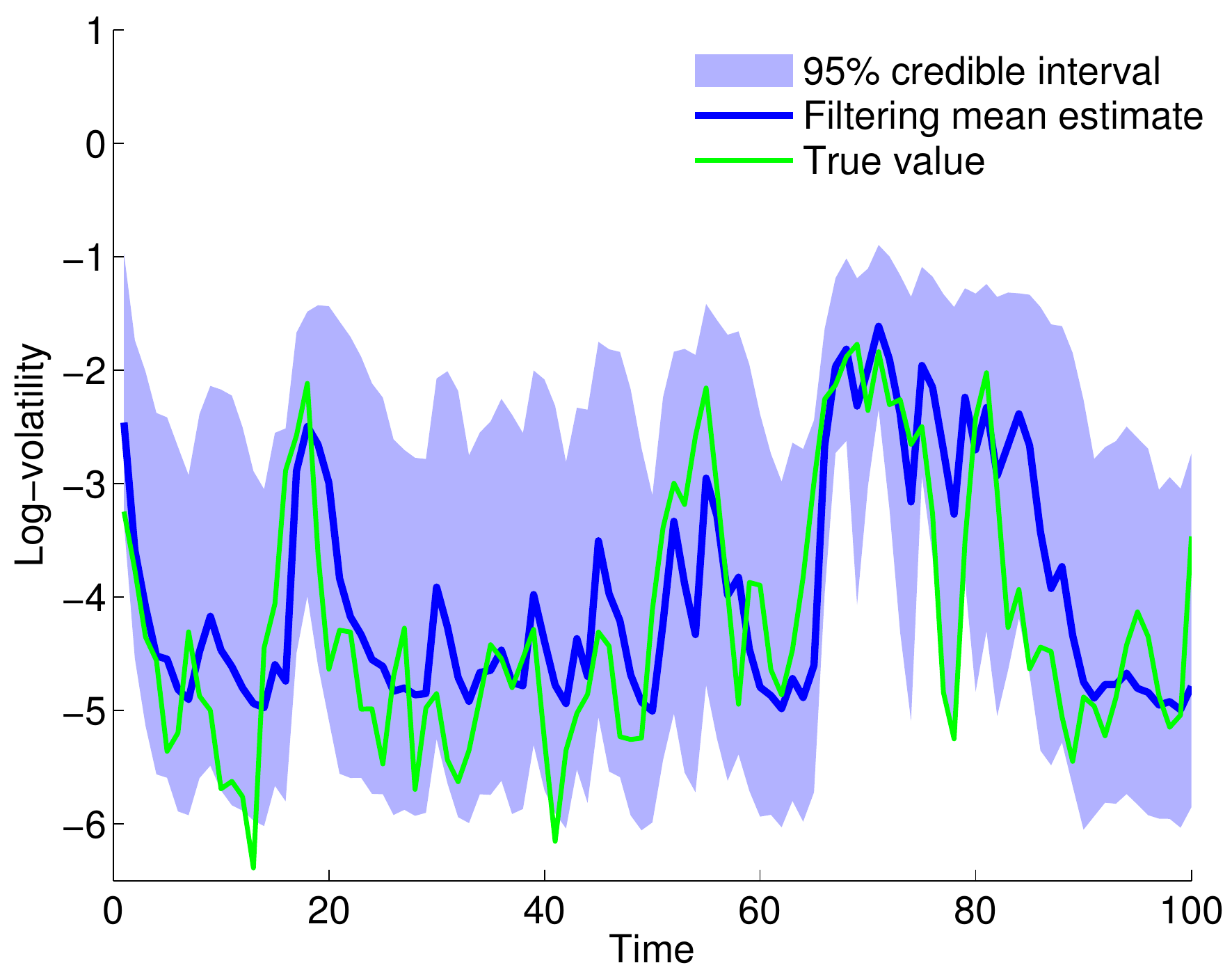}}
\subfigure[Smoothing]{\includegraphics[width=.48\textwidth]{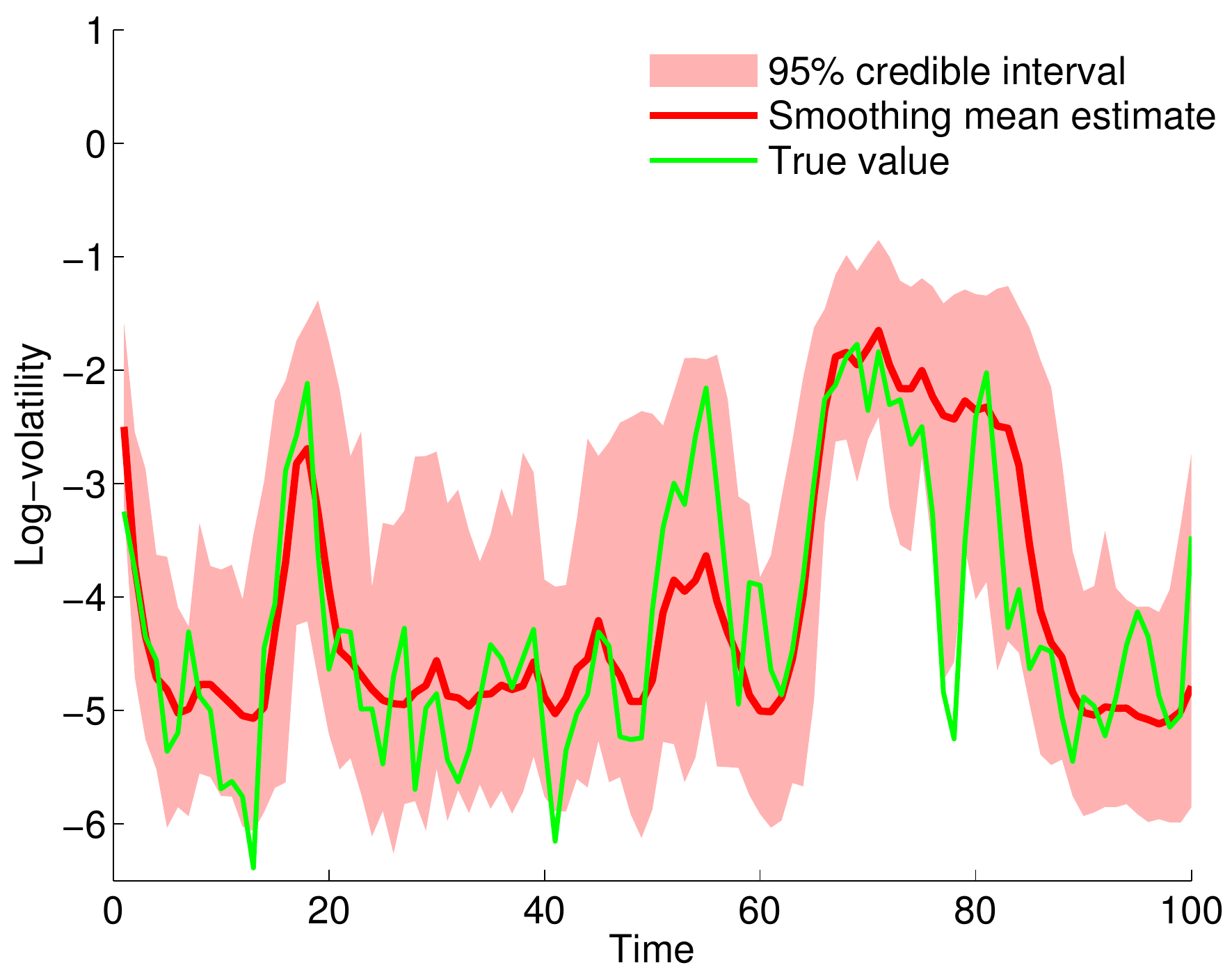}}
\end{center}
\caption{SMC: (a) Filtering and (b) smoothing estimates and credible intervals for the switching stochastic volatility model.}
\label{fig:volatility_estimates}
\end{figure}

\begin{figure}
\begin{center}
\includegraphics[width=10cm]{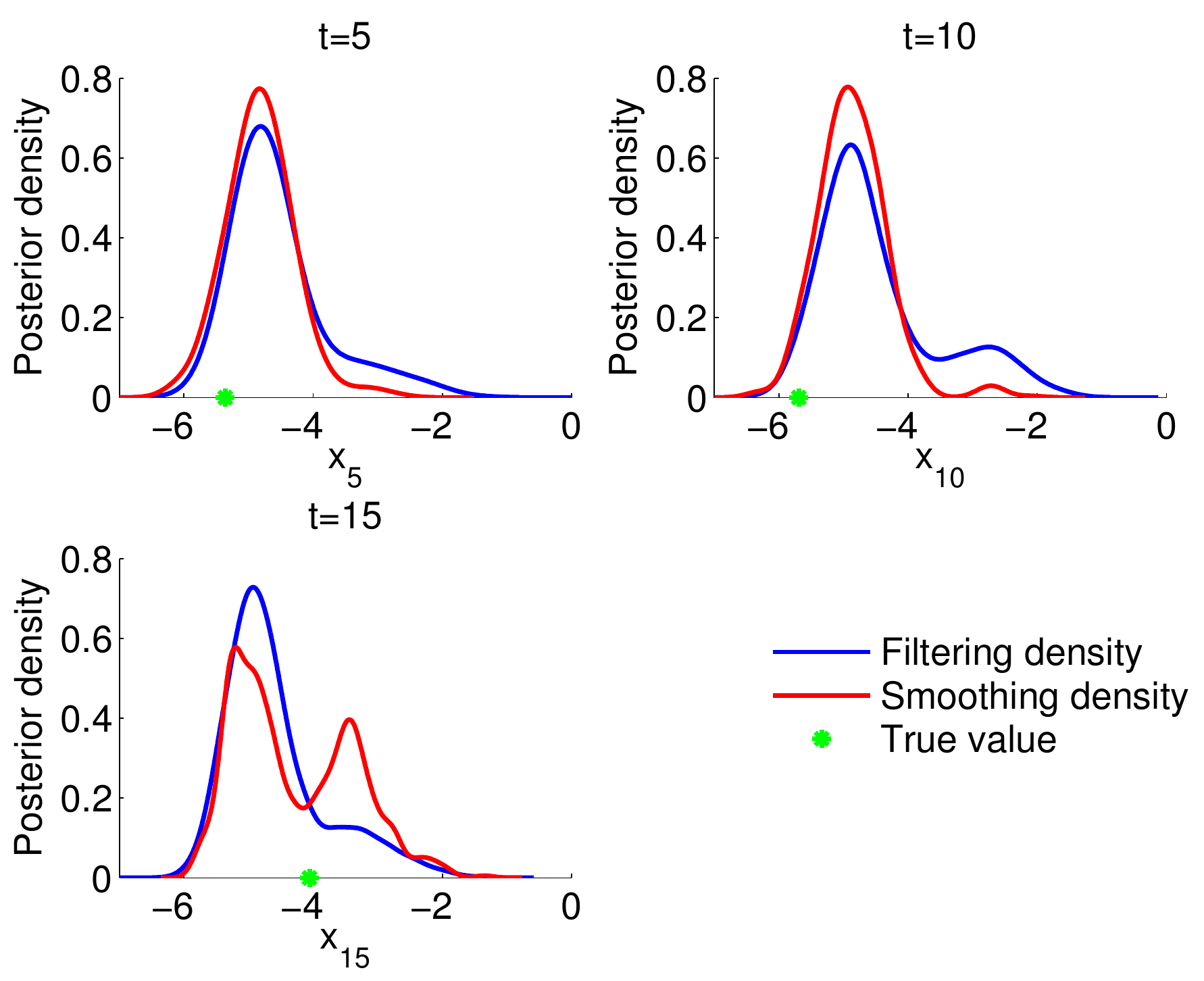}
\end{center}
\caption{SMC: Kernel density estimates for the marginal posteriors of $X_t|Y_{1:t}$ and  $X_t|Y_{1:t_\text{max}}$ for $t=5,10,15$.}
\label{fig:volatility_kde}
\end{figure}

Plots of these summary statistics and kernel density estimates are given in Figures~\ref{fig:volatility_estimates} and~\ref{fig:volatility_kde}.

\subsection{Bayesian inference with particle independent Metropolis-Hastings}

The SMC algorithm can also be used as a proposal distribution within an MCMC algorithm to provide MCMC samples from the posterior distribution, as described in Section~\ref{sec:pmcmc}. This can be done with \Biips\ with the functions \code{biips_pimh_init}, \code{biips_pimh_update} and \code{biips_pimh_samples}. The first function creates a PIMH object; The second one runs burn-in iterations; The third one runs PIMH iterations and returns samples.

\begin{lstlisting}[style=matbiips]
n_burn = 2000; n_iter = 10000; thin = 1; n_part = 50;
obj_pimh = biips_pimh_init(model, variables);
obj_pimh = biips_pimh_update(obj_pimh, n_burn, n_part); % Burn-in iterations
[obj_pimh, out_pimh, log_marg_like_pimh] = biips_pimh_samples(obj_pimh,...
    n_iter, n_part, 'thin', thin); % Return samples

summ_pimh = biips_summary(out_pimh, 'probs', [.025, .975]);
x_pimh_mean = summ_pimh.x.mean;
x_pimh_quant = summ_pimh.x.quant;
\end{lstlisting}

\begin{lstlisting}[style=rbiips]
n_burn <- 2000; n_iter <- 10000; thin <- 1; n_part <- 50
obj_pimh <- biips_pimh_init(model, variables)
biips_pimh_update(obj_pimh, n_burn, n_part) # Burn-in iterations
out_pimh <- biips_pimh_samples(obj_pimh, n_iter, n_part,
                               thin=thin) # Return samples

summ_pimh <- biips_summary(out_pimh, probs=c(.025, .975))
x_pimh_mean <- summ_pimh$x$mean
x_pimh_quant <- summ_pimh$x$quant
\end{lstlisting}

\begin{figure}
\begin{center}
\subfigure[Estimates]{\includegraphics[width=.48\textwidth]{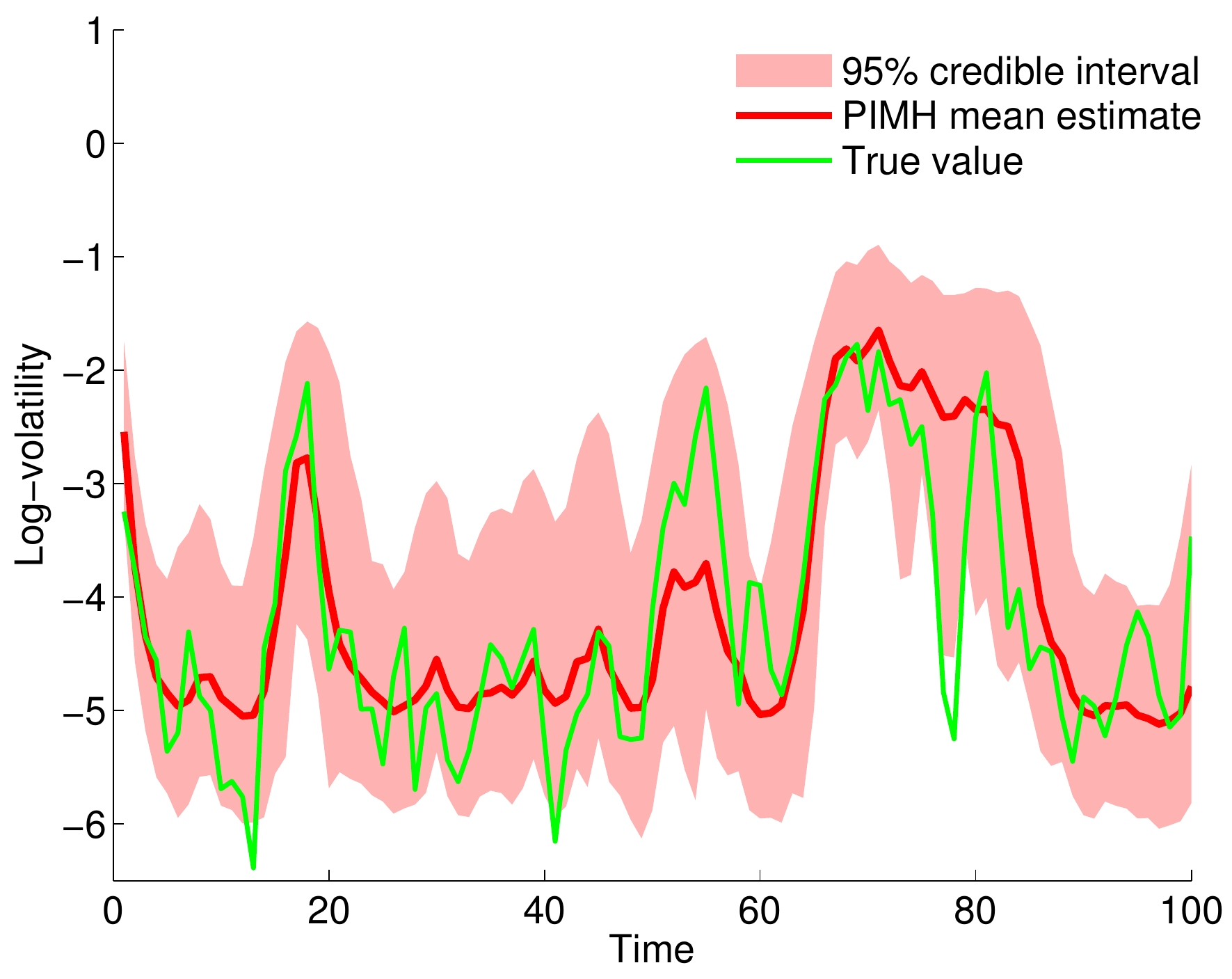}}
\subfigure[Posterior marginals]{\includegraphics[width=.48\textwidth]{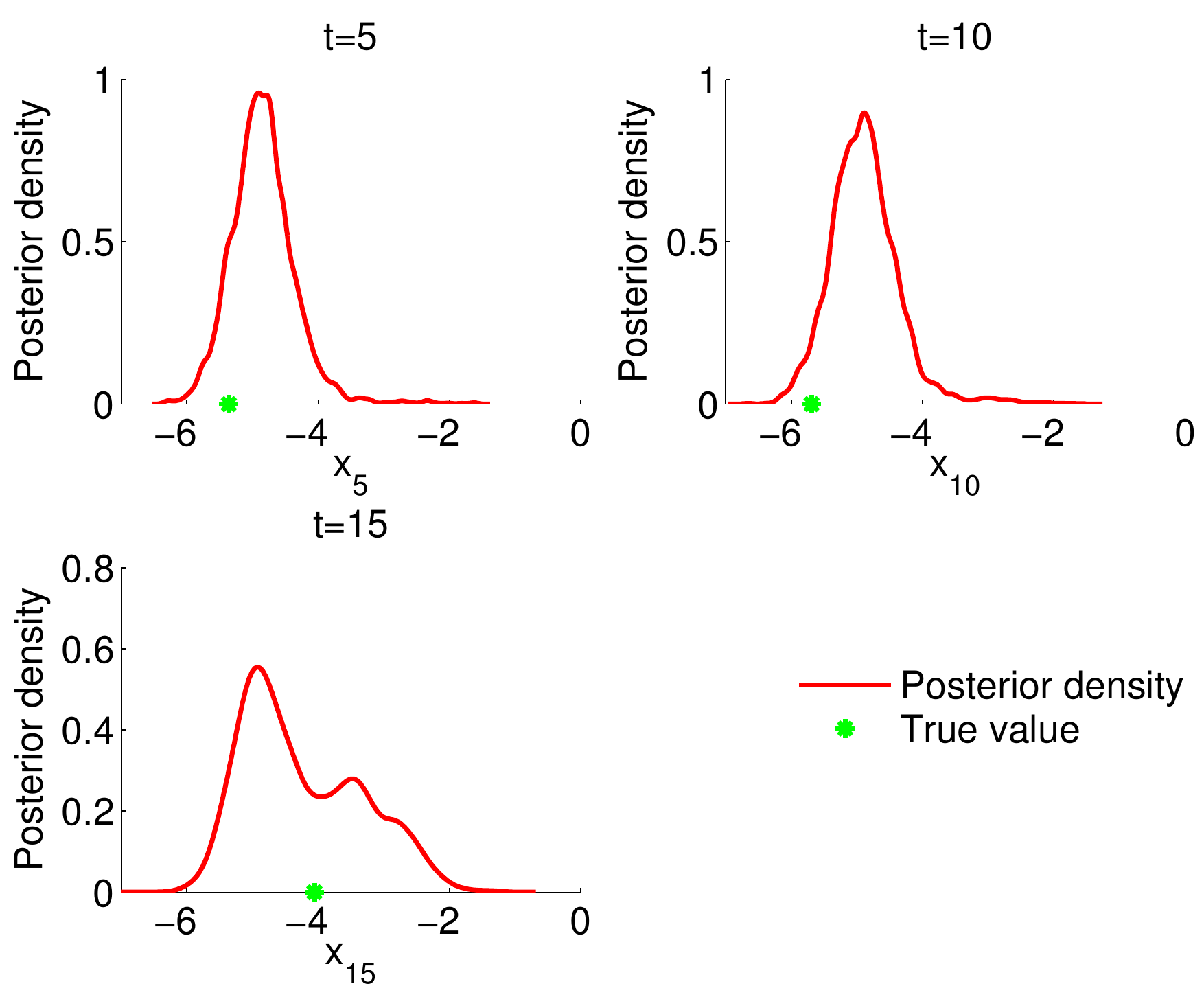}}
\end{center}
\caption{PIMH: (a) Smoothing estimates and credible intervals for the switching stochastic volatility model. (b) Posterior marginals $p(x_t|y_{1:t_{\max}})$ for $t=5,10,15$.}
\label{fig:volatility_pimh}
\end{figure}

Posterior means, credible intervals and some marginal posteriors are reported in Figure~\ref{fig:volatility_pimh}.

\subsection{Sensitivity analysis with SMC}

We now consider evaluating the sensitivity of the model with respect to the model parameters $\alpha_1$ and $\alpha_2$. For a grid of values of those parameters, we report the estimated logarithm of the marginal likelihood $p(y_{1:t_{\max}})$ using a sequential Monte Carlo algorithm.

\begin{lstlisting}[style=matbiips]
n_part = 50;
param_names = {'alpha'};
[A, B] = meshgrid(-5:.2:2, -5:.2:2);
param_values = {[A(:), B(:)]'};

out_sens = biips_smc_sensitivity(model, param_names, param_values, n_part);
\end{lstlisting}

\begin{lstlisting}[style=rbiips]
n_part <- 50
range <- seq(-5,2,.2)
A <- rep(range, times=length(range))
B <- rep(range, each=length(range))
param_values <- list('alpha'=rbind(A, B))

out_sens <- biips_smc_sensitivity(model, param_values, n_part)
\end{lstlisting}

\begin{figure}
\begin{center}
\includegraphics[width=10cm]{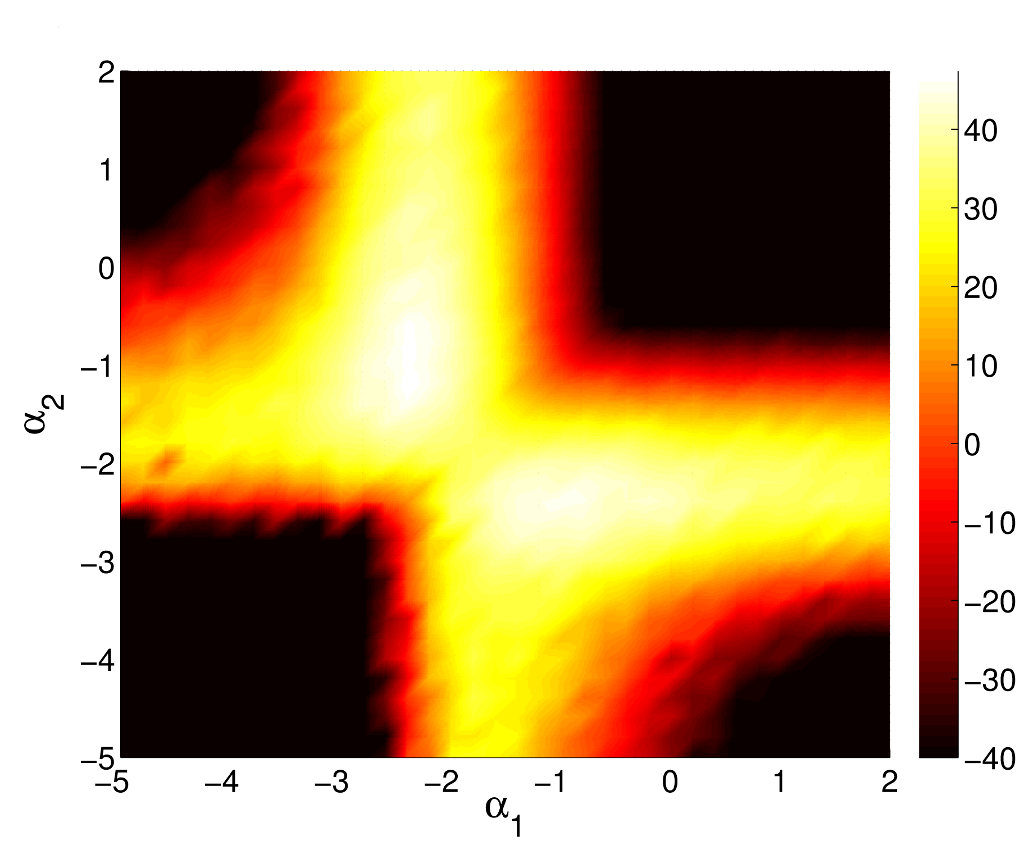}
\end{center}
\caption{Sensitivity: Estimates of the marginal log-likelihood provided by SMC for different values of $\alpha_1$ and $\alpha_2$, the other values of the parameters being held fixed. True values are $\alpha_1=-2.5$ and $\alpha_2=-1$.}
\label{fig:volatility_sensitivity}
\end{figure}

The results are reported in Figure~\ref{fig:volatility_sensitivity}.

\newpage

\subsection{Bayesian inference with unknown parameters with the particle marginal Metropolis-Hastings}

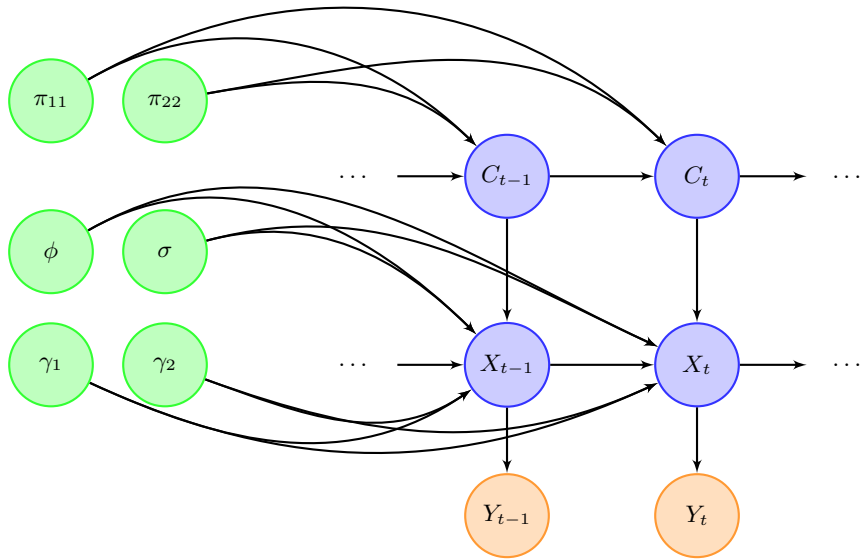
\begin{figure}
\begin{center}
\begin{tikzpicture}[node distance=2cm,auto,>=latex',minimum size=1.1cm]
\node (c_t-2)   [] {\ldots};
\node (c_t-1)   [state,right of=c_t-2] {$C_{t-1}$};
\node (c_t)   [state,right of=c_t-1,node distance=2.5cm] {$C_t$};
\node (c_t+1)   [right of=c_t] {\ldots};
\node (X_t-2)   [below of=c_t-2,node distance=2.5cm] {\ldots};
\node (X_t-1)   [state,right of=X_t-2] {$X_{t-1}$};
\node (X_t)   [state,right of=X_t-1,node distance=2.5cm] {$X_t$};
\node (X_t+1)   [right of=X_t] {\ldots};
 \node (Y_t)   [measurement,below of=X_t] {$Y_t$};
 \node (Y_t-1)   [measurement,below of=X_t-1] {$Y_{t-1}$};

   \node (gamma_2)   [param,left of=X_t-2,node distance=2.5cm] {$\gamma_2$};
 \node (gamma_1)   [param,left of=gamma_2,node distance=1.5cm] {$\gamma_1$};
   \node (phi)   [param,above of=gamma_1,node distance=1.5cm] {$\phi$};
     \node (sigma)   [param,right of=phi,node distance=1.5cm] {$\sigma$};

   \node (pi_22)   [param,above of=sigma,node distance=2cm] {$\pi_{22}$};
   \node (pi_11)   [param,left of=pi_22,node distance=1.5cm] {$\pi_{11}$};

 \path[->] (c_t-1) edge[thick] (c_t);
 \path[->] (c_t-2) edge[thick] (c_t-1);
 \path[->] (c_t) edge[thick] (c_t+1);
 \path[->] (X_t-1) edge[thick] (X_t);
 \path[->] (X_t-2) edge[thick] (X_t-1);
 \path[->] (X_t) edge[thick] (X_t+1);
  \path[->] (c_t) edge[thick] (X_t);
 \path[->] (c_t-1) edge[thick] (X_t-1);
 \path[->] (X_t) edge[thick] (Y_t);
 \path[->] (X_t-1) edge[thick] (Y_t-1);
 \path[->] (gamma_1) edge[thick,out=-25,in=-145] (X_t-1);
 \path[->] (gamma_1) edge[thick,out=-25,in=-155] (X_t);
 \path[->] (gamma_2) edge[thick,out=-20,in=-145] (X_t-1);
 \path[->] (gamma_2) edge[thick,out=-20,in=-155] (X_t);
  \path[->] (phi) edge[thick,out=30] (X_t-1);
 \path[->] (phi) edge[thick,out=30,in=155] (X_t);
   \path[->] (sigma) edge[thick,out=15] (X_t-1);
 \path[->] (sigma) edge[thick,out=15,in=155] (X_t);

    \path[->] (pi_11) edge[thick,out=30] (c_t-1);
    \path[->] (pi_11) edge[thick,out=30] (c_t);
        \path[->] (pi_22) edge[thick,out=10] (c_t-1);
    \path[->] (pi_22) edge[thick,out=10] (c_t);
 \end{tikzpicture}
 \caption{Graphical representation of the full switching volatility model defined by Equations~\eqref{eq:volatilitymodel} and \eqref{eq:volatilitymodel2}  as a directed acyclic graph. Blue and green nodes correspond to unobserved variables, orange nodes to observed variables. In the particle marginal Metropolis-Hastings algorithm, green nodes correspond to variables sampled using a Metropolis-Hastings proposal, whereas blue nodes correspond to variables sampled using a SMC algorithm.}
 \label{fig:graphparam}
 \end{center}
 \end{figure}

So far, we have assumed that the parameters $\alpha$, $\pi$, $\phi$ and $\tau$ were fixed and known. We now consider that these variables have to be estimated as well.
We consider the following prior on the parameters~\citep{Carvalho2007}:
\begin{equation}
\begin{aligned}
\alpha_1&=\gamma_1\\
\alpha_2&=\gamma_1+\gamma_2\\
\gamma_1&\sim \Norm (0,100)\\
\gamma_2&\sim \TNorm_{(0,+\infty)} (0,100)
\end{aligned}
\qquad\qquad
\begin{aligned}
\frac{1}{\sigma^2}&\sim \Gam(2.001,1)\\
\phi&\sim\TNorm_{(-1,1)} (0,100)\\
\pi_{11}&\sim \Beta(10,1)\\
\pi_{22}&\sim \Beta(10,1)
\end{aligned}
\label{eq:volatilitymodel2}
\end{equation}
where $\Gam(a,b)$ is the standard Gamma distribution of scale $a>0$ and rate $b>0$, $\TNorm_{(a,b)} (\mu,\sigma^2)$ is the truncated normal distribution of mean $\mu\in\mathbb R$ and variance $\sigma^2>0$ with support $[a,b]$, $-\infty<a<b<\infty$ and $\Beta(a,b)$ is the standard beta distribution with parameters $a>0$ and $b>0$. Note that the prior on $\phi$ is essentially uniform. Figure~\ref{fig:graphparam} shows the representation of the full statistical model as a directed acyclic graph.

The Listing \ref{listing:sswbugs2} provides the transcription of the statistical model defined by Equations~\eqref{eq:volatilitymodel} and \eqref{eq:volatilitymodel2} in \BUGS\ language.

\begin{lstlisting}[style=bugs,caption=Switching stochastic volatility model with unknown parameters in BUGS language,label=listing:sswbugs2,float=ht!]
model
{
  gamma[1] ~ dnorm(0, 1/100)
  gamma[2] ~ dnorm(0, 1/100) T(0,)
  alpha[1] <- gamma[1]
  alpha[2] <- gamma[1] + gamma[2]
  phi ~ dnorm(0, 1/100) T(-1,1)
  tau ~ dgamma(2.001, 1)
  sigma <- 1/sqrt(tau)
  pi[1,1] ~ dbeta(10, 1)
  pi[1,2] <- 1 - pi[1,1]
  pi[2,2] ~ dbeta(10, 1)
  pi[2,1] <- 1 - pi[2,2]

  c[1] ~ dcat(pi[1,])
  mu[1] <- alpha[1] * (c[1] == 1) + alpha[2] * (c[1] == 2)
  x[1] ~ dnorm(mu[1], 1/sigma^2) T(-500,500)
  prec_y[1] <- exp(-x[1])
  y[1] ~ dnorm(0, prec_y[1])
  for (t in 2:t_max)
  {
    c[t] ~ dcat(ifelse(c[t-1] == 1, pi[1,], pi[2,]))
    mu[t] <- alpha[1] * (c[t] == 1) + alpha[2] * (c[t] == 2) + phi * x[t-1]
    x[t] ~ dnorm(mu[t], 1/sigma^2) T(-500,500)
    prec_y[t] <- exp(-x[t])
    y[t] ~ dnorm(0, prec_y[t])
  }
}
\end{lstlisting}

The user can then load the model and run a PMMH sampler to approximate the joint distribution of $\theta=(\alpha_1,\alpha_2,\sigma,\pi_{11},\pi_{22},\phi)$ and $(X_1,C_1,\ldots,X_{t_{\max}},C_{t_{\max}})$ given the data $(Y_1,\ldots,Y_{t_{\max}})$. The function \code{biips_pmmh_init} creates a PMMH object. The input \code{param_names} contains the names of the variables to be updated using a Metropolis-Hastings proposal, here $(\gamma_1,\gamma_2,\tau=\frac{1}{\sigma^2},\phi,\pi_{11},\pi_{22})$. Other variables are updated using a SMC algorithm. The input \code{latent_names} specifies the other variables for which we want to obtain posterior samples. The function \code{biips_pmmh_update} runs a PMMH with adaptation and burn-in iterations. During the adaptation phase, it learns the parameters of the proposal distribution $\nu$ in Algorithm \ref{algo:pmmh}.

\begin{lstlisting}[style=matbiips]
sigma_true = .4; alpha_true = [-2.5; -1]; phi_true = .5;
pi11 = .9; pi22 = .9; pi_true = [pi11, 1 - pi11; 1 - pi22, pi22];
data = struct('t_max', t_max, 'sigma_true', sigma_true,...
    'alpha_true', alpha_true, 'phi_true', phi_true, 'pi_true', pi_true);
model_file = 'switch_stoch_volatility_param.bug';
model = biips_model(model_file, data, 'sample_data', sample_data);
data = model.data;

n_burn = 2000; n_iter = 40000; thin = 10; n_part = 50;
param_names = {'gamma[1]', 'gamma[2]', 'phi', 'tau', 'pi[1,1]', 'pi[2,2]'};
latent_names = {'x', 'alpha[1]', 'alpha[2]', 'sigma'};

inits = {-1, 1, .5,5, .8, .8};
obj_pmmh = biips_pmmh_init(model, param_names,...
    'inits', inits, 'latent_names', latent_names);
obj_pmmh = biips_pmmh_update(obj_pmmh, n_burn, n_part);
[obj_pmmh, out_pmmh, log_marg_like_pen, log_marg_like] =...
    biips_pmmh_samples(obj_pmmh, n_iter, n_part, 'thin', thin);
\end{lstlisting}

\begin{lstlisting}[style=rbiips]
sigma_true <- .4; alpha_true <- c(-2.5, -1); phi_true <- .5
pi11 <- .9; pi22 <- .9
pi_true <- matrix(c(pi11, 1-pi11, 1-pi22, pi22), nrow=2, byrow=TRUE)
data <- list(t_max=t_max, sigma_true=sigma_true,
             alpha_true=alpha_true, phi_true=phi_true, pi_true=pi_true)
model_file <- 'switch_stoch_volatility_param.bug'
model <- biips_model(model_file, data, sample_data=sample_data)
data <- model$data()

n_burn <- 2000; n_iter <- 40000; thin <- 10; n_part <- 50
param_names <- c('gamma[1]', 'gamma[2]', 'phi', 'tau', 'pi[1,1]', 'pi[2,2]')
latent_names <- c('x', 'alpha[1]', 'alpha[2]', 'sigma')
inits <- list(-1, 1, .5, 5, .8, .8)
obj_pmmh <- biips_pmmh_init(model, param_names, inits=inits,
                            latent_names=latent_names)
biips_pmmh_update(obj_pmmh, n_burn, n_part)
out_pmmh <- biips_pmmh_samples(obj_pmmh, n_iter, n_part, thin=thin)
\end{lstlisting}

\begin{figure}[ht!]
\begin{center}
\includegraphics[width=.32\textwidth]{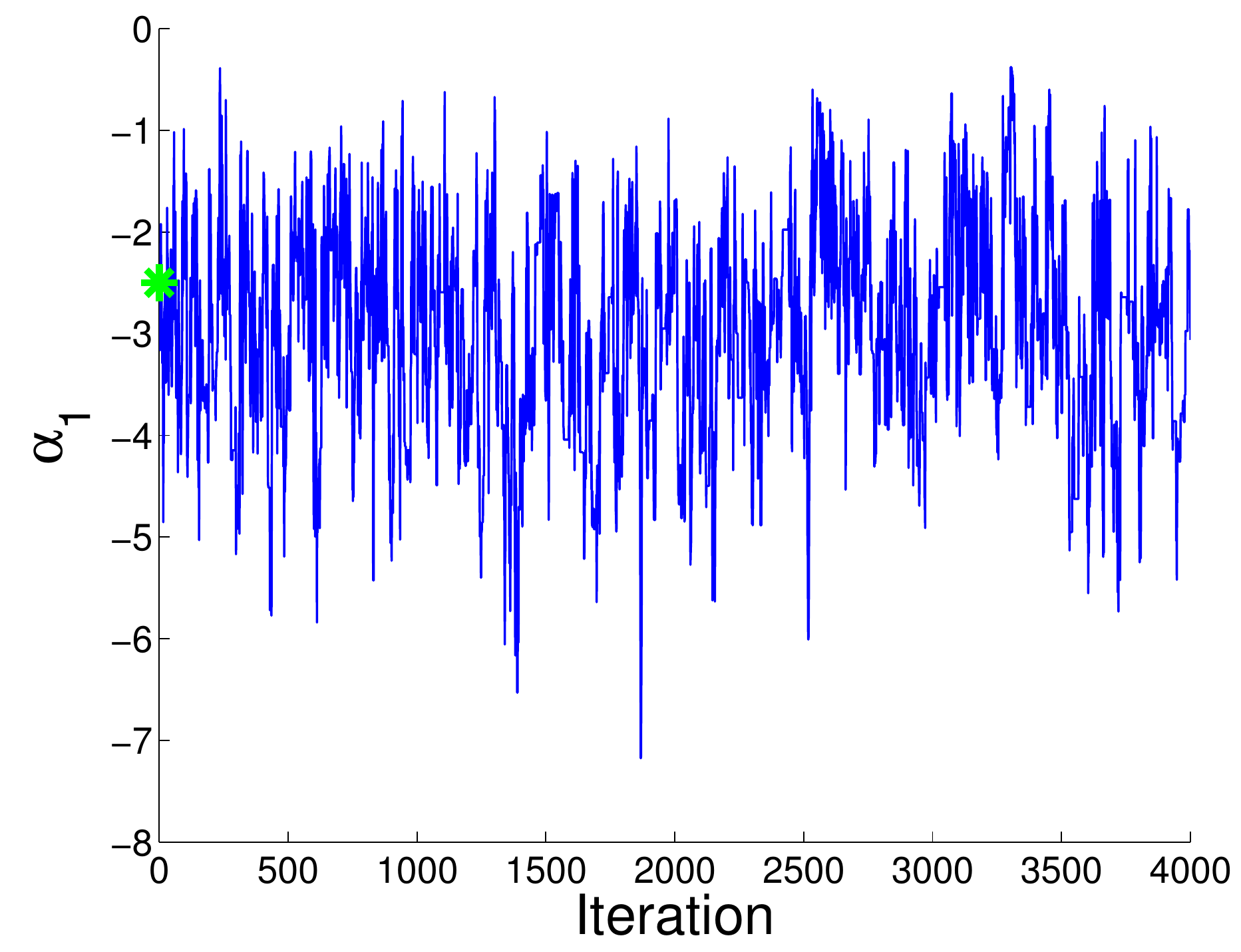}
\includegraphics[width=.32\textwidth]{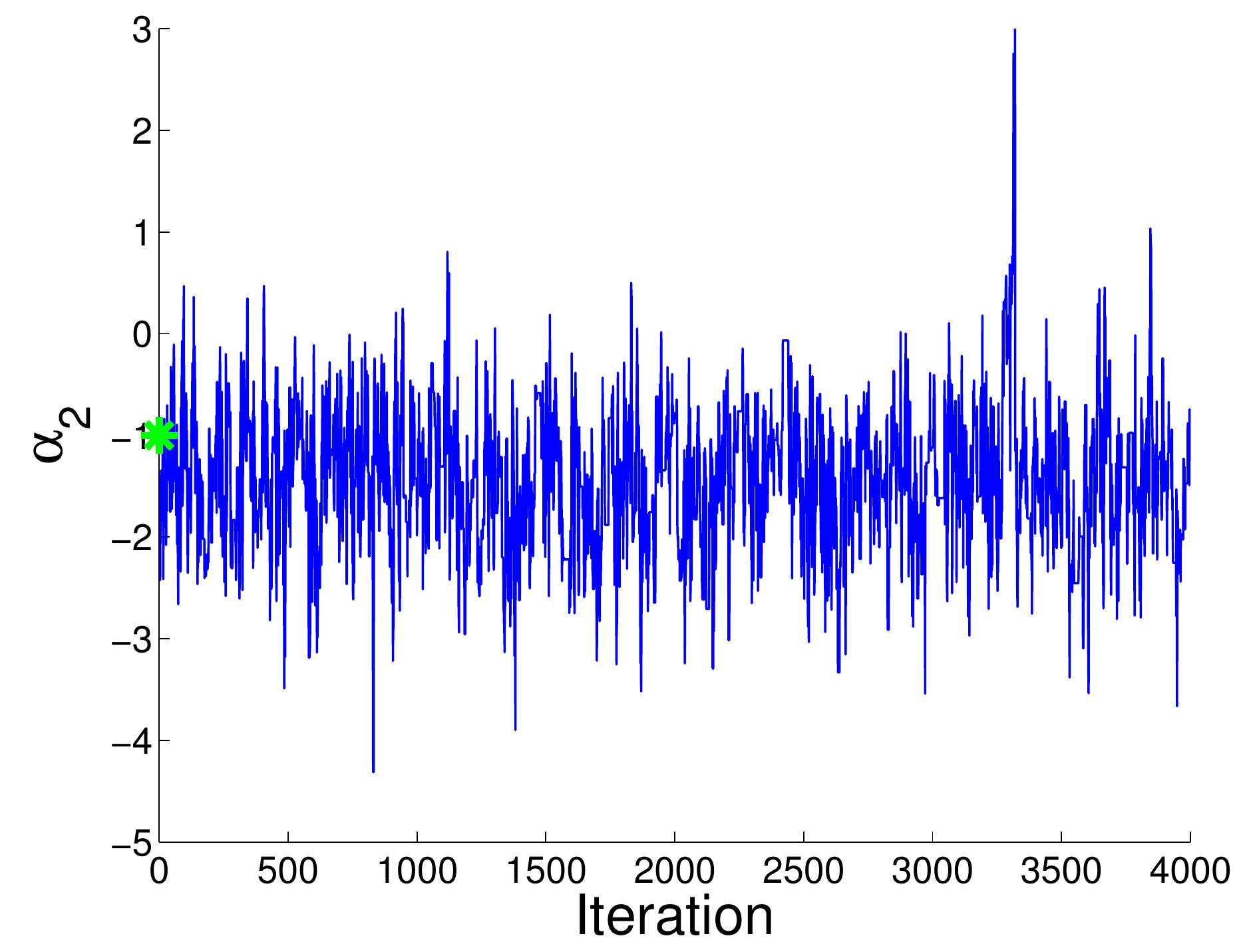}
\includegraphics[width=.32\textwidth]{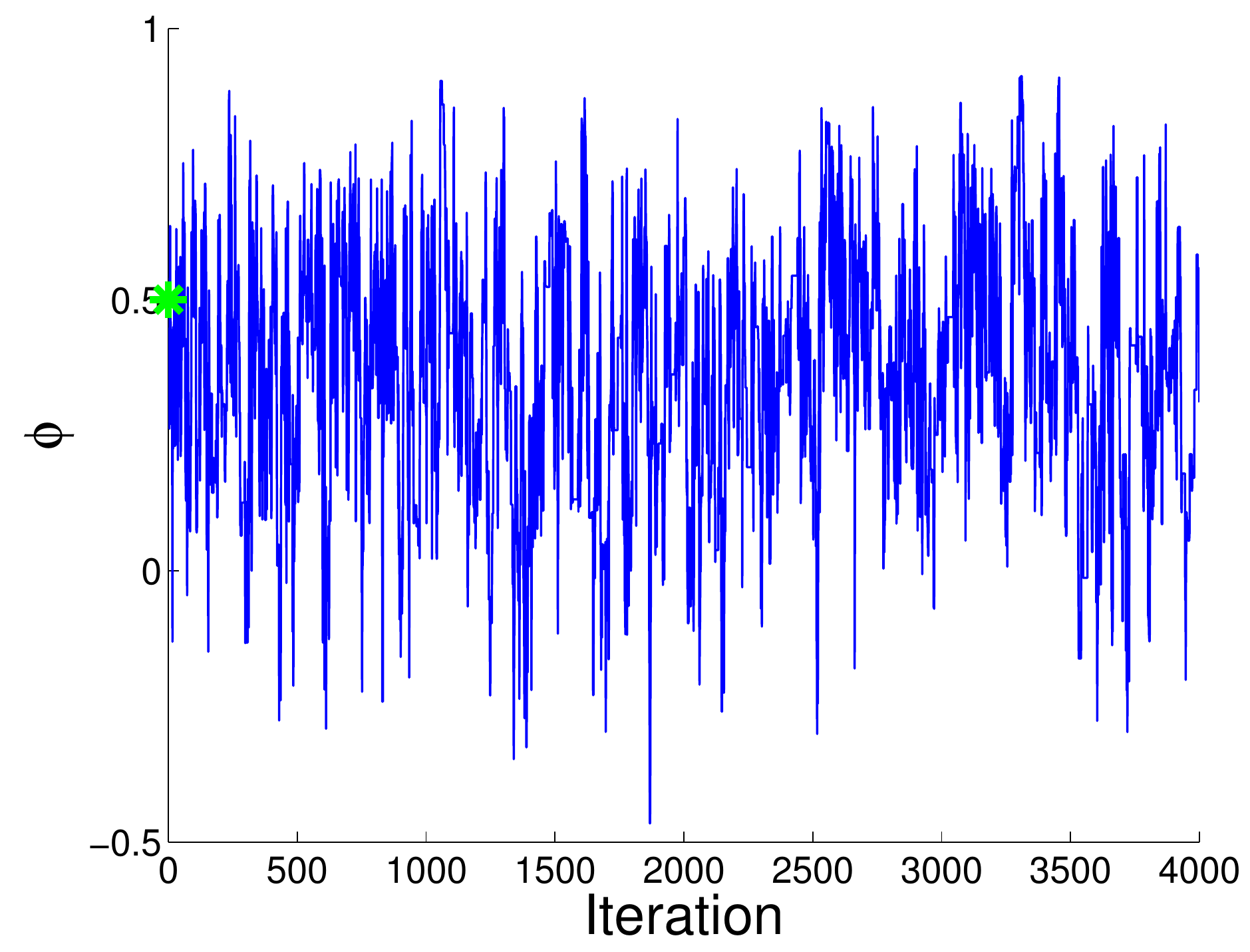}
\subfigure[$\alpha_1$]{ \includegraphics[width=.32\textwidth]{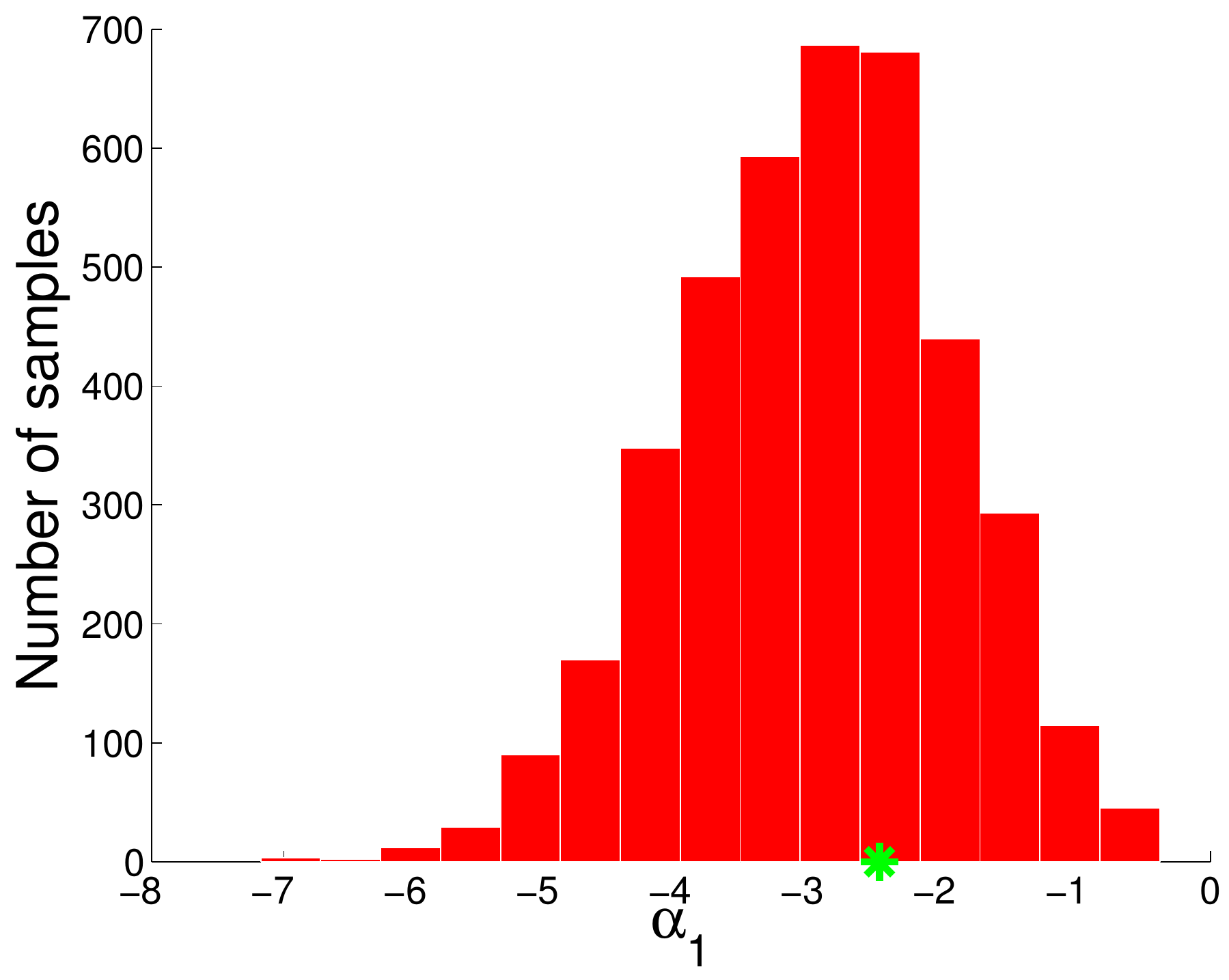}}
\subfigure[$\alpha_2$]{\includegraphics[width=.32\textwidth]{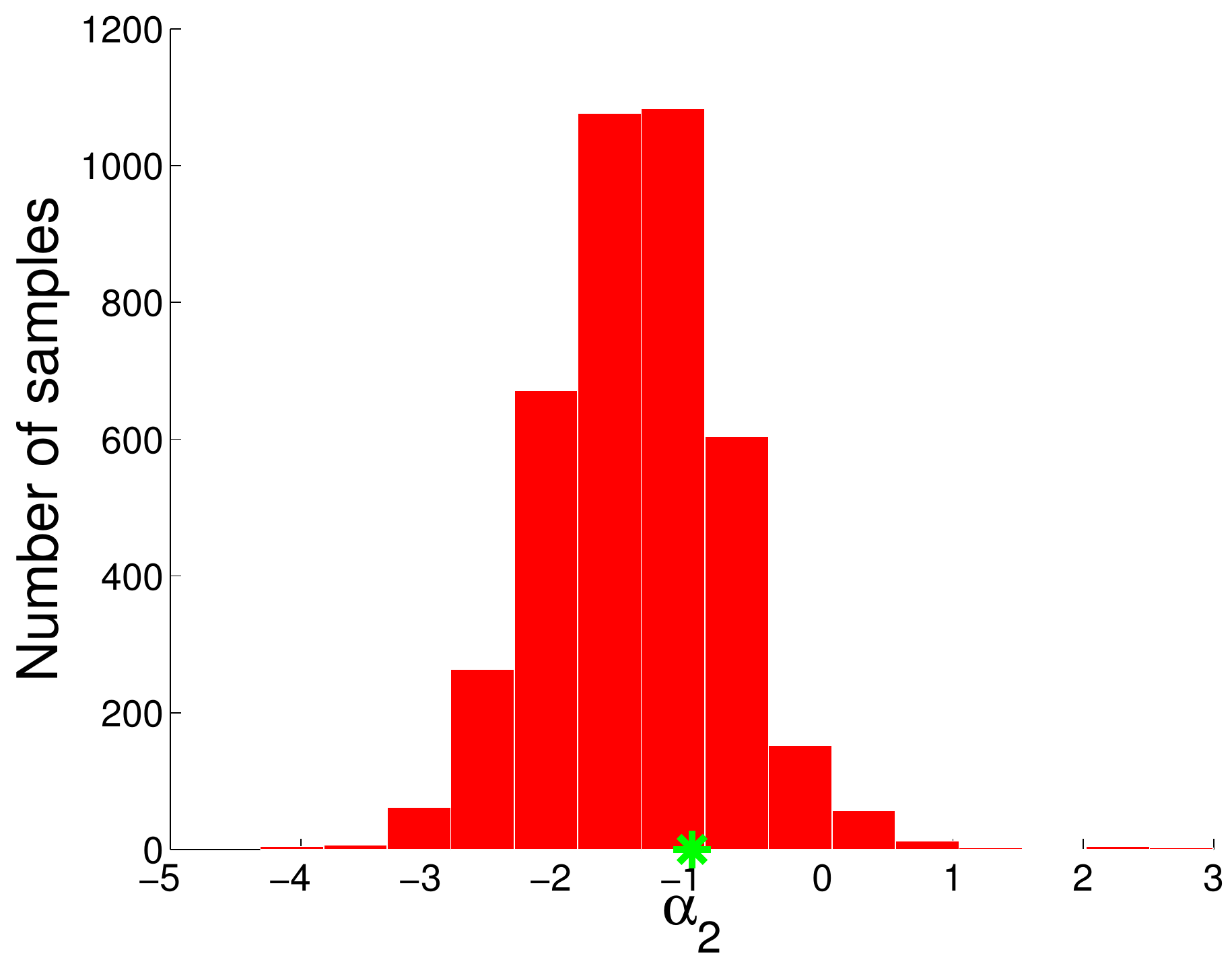}}
\subfigure[$\phi$]{\includegraphics[width=.32\textwidth]{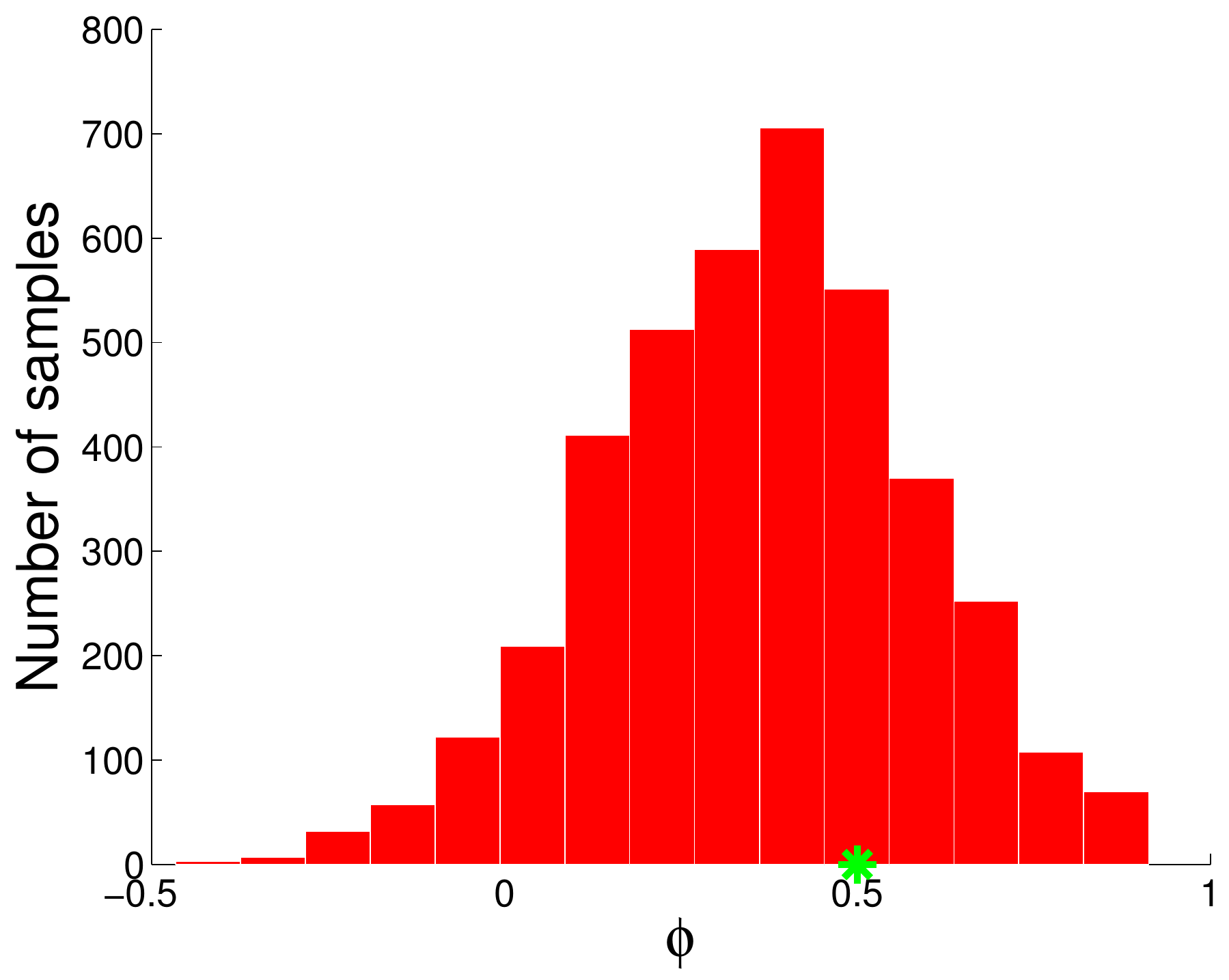}}
\includegraphics[width=.32\textwidth]{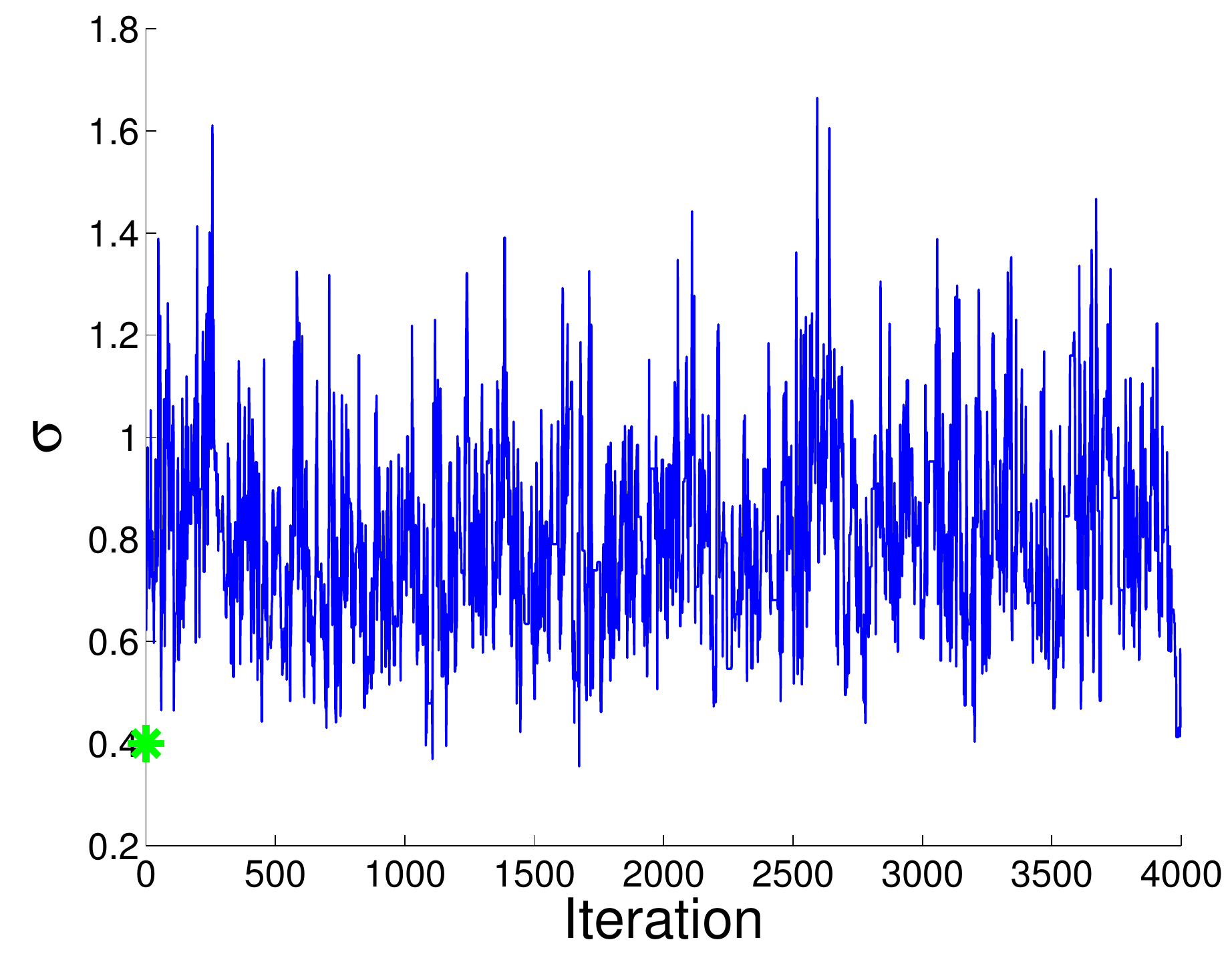}
\includegraphics[width=.32\textwidth]{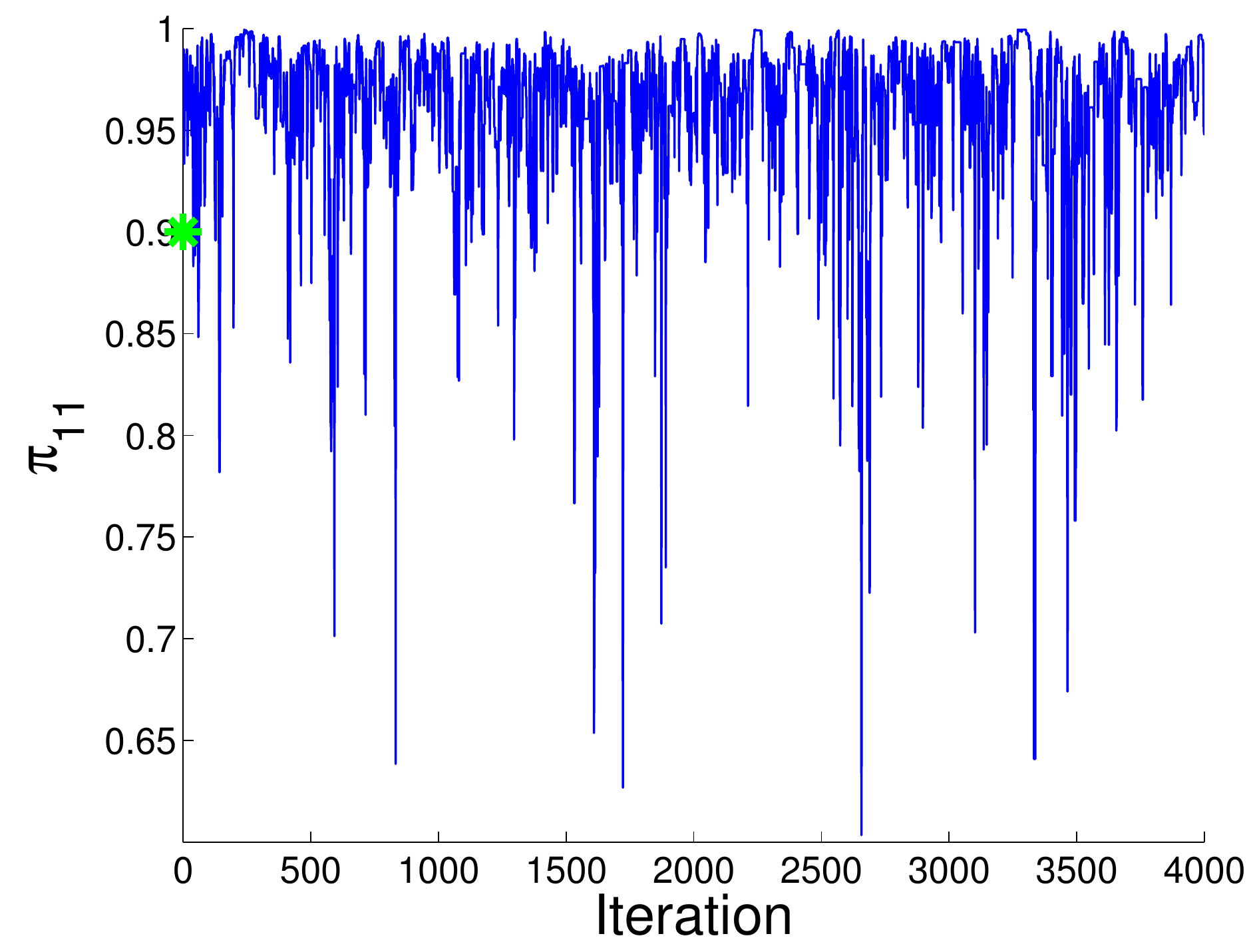}
\includegraphics[width=.32\textwidth]{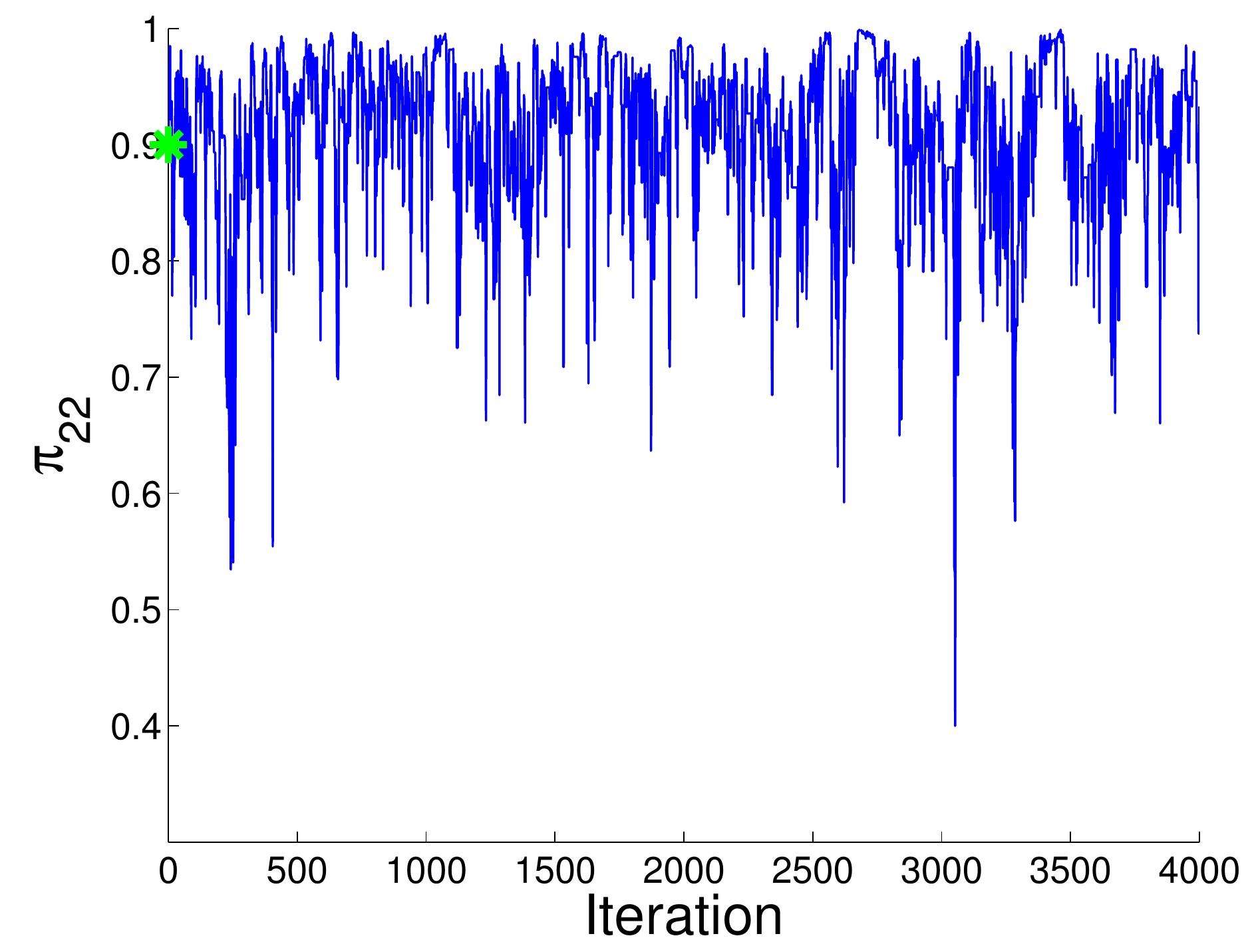}
\subfigure[$\sigma$]{\includegraphics[width=.32\textwidth]{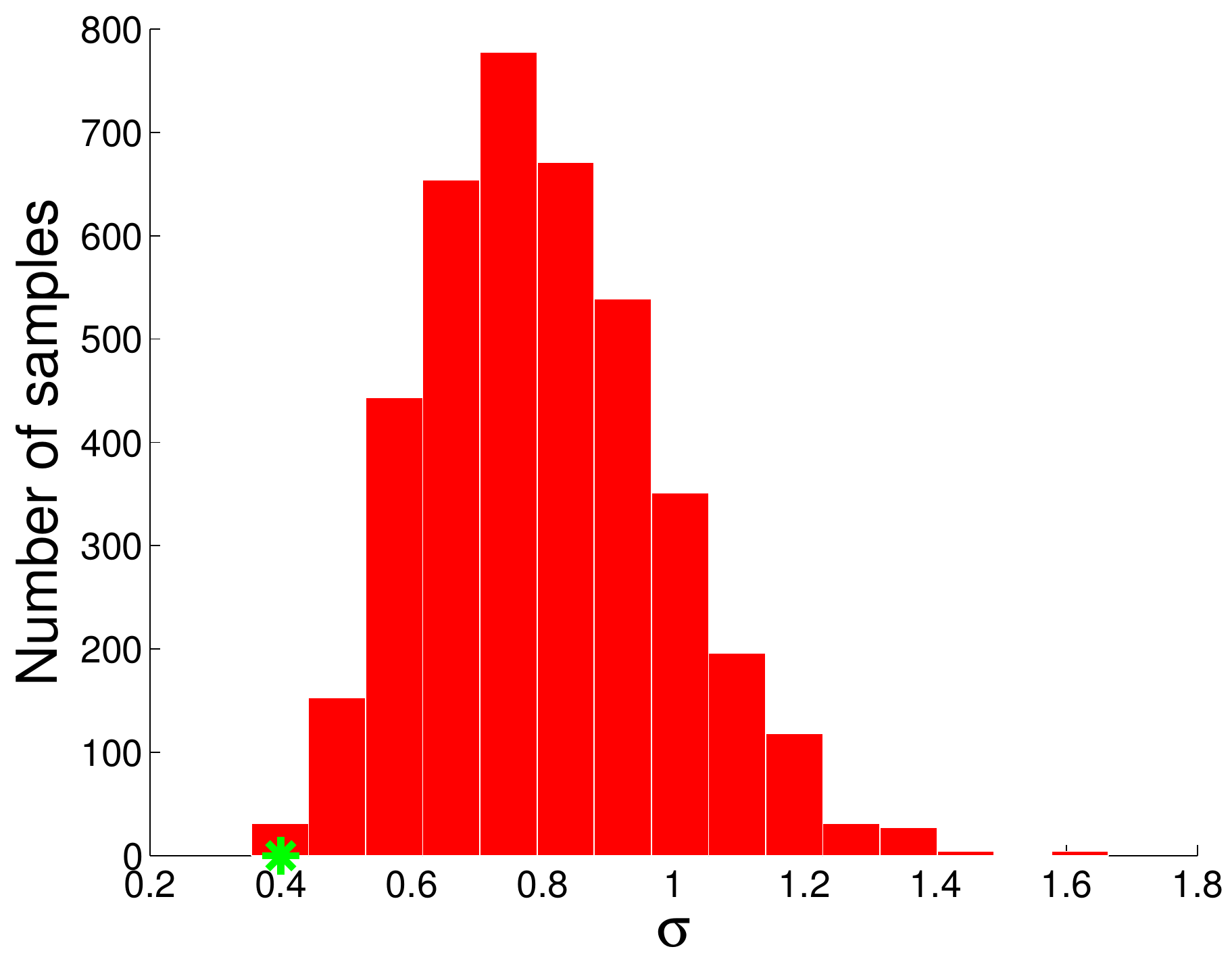}}
\subfigure[$\pi_{11}$]{\includegraphics[width=.32\textwidth]{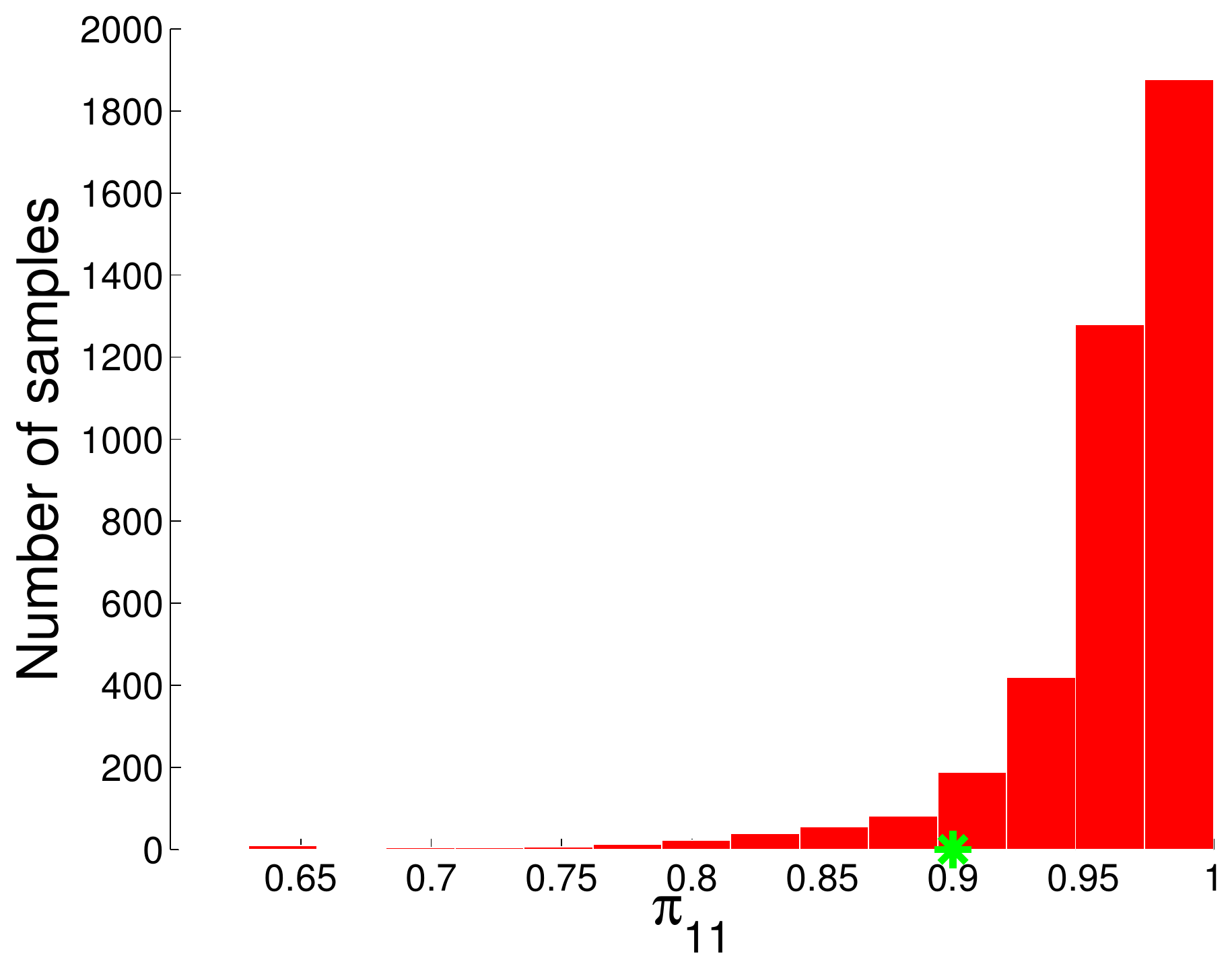}}
\subfigure[$\pi_{22}$]{\includegraphics[width=.32\textwidth]{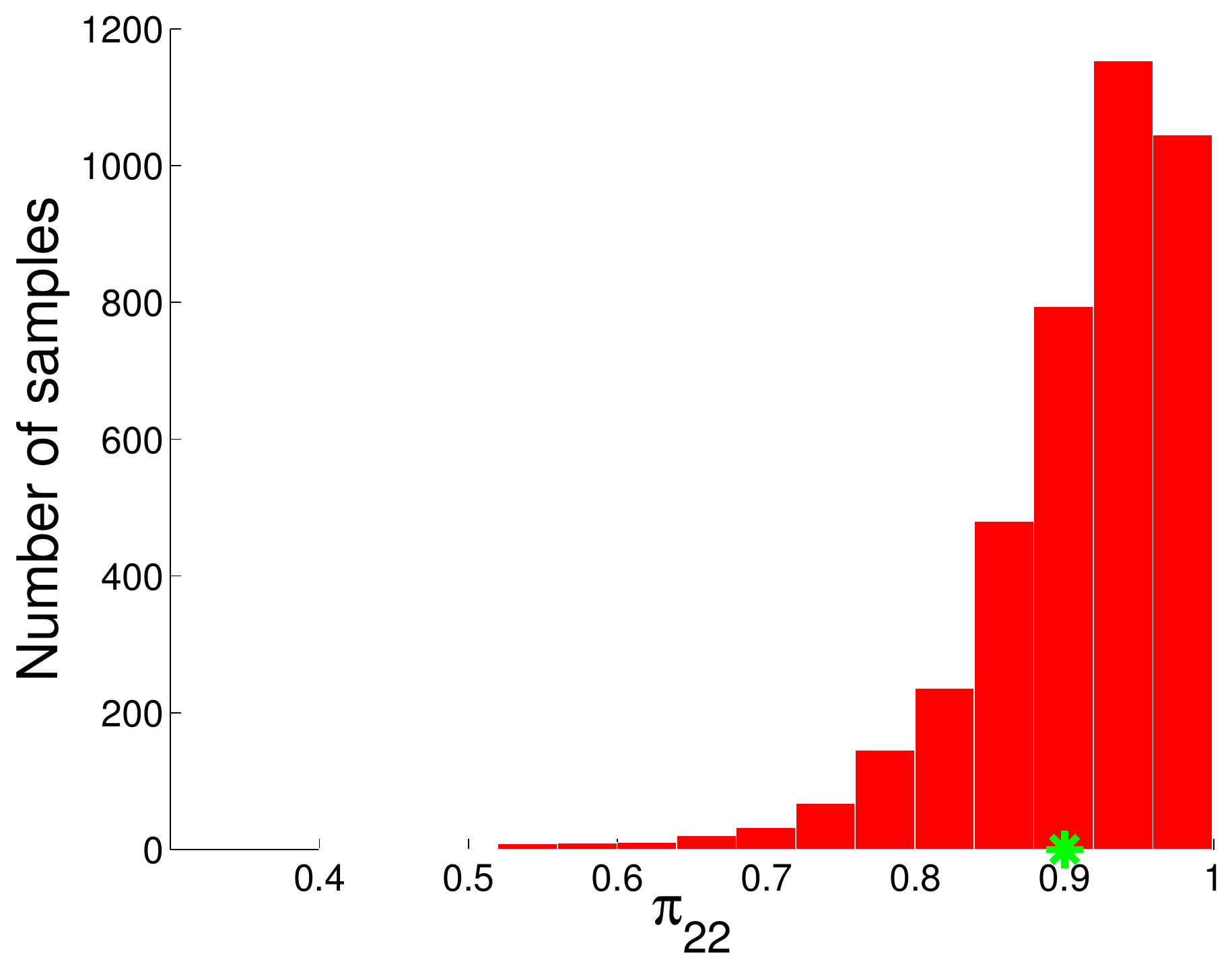}}
\end{center}
\caption{PMMH: Posterior samples traces (top figures)  and histograms (bottom figures) of the parameters (a) $\alpha_1$, (b) $\alpha_2$, (c) $\phi$, (d) $\sigma$ (e) $\pi_{11}$ and (f) $\pi_{22}$. True values are represented by a green star.}
\label{fig:volatility_pmmh_param}
\end{figure}

\begin{figure}
\begin{center}
\subfigure[Estimates]{\includegraphics[width=.48\textwidth]{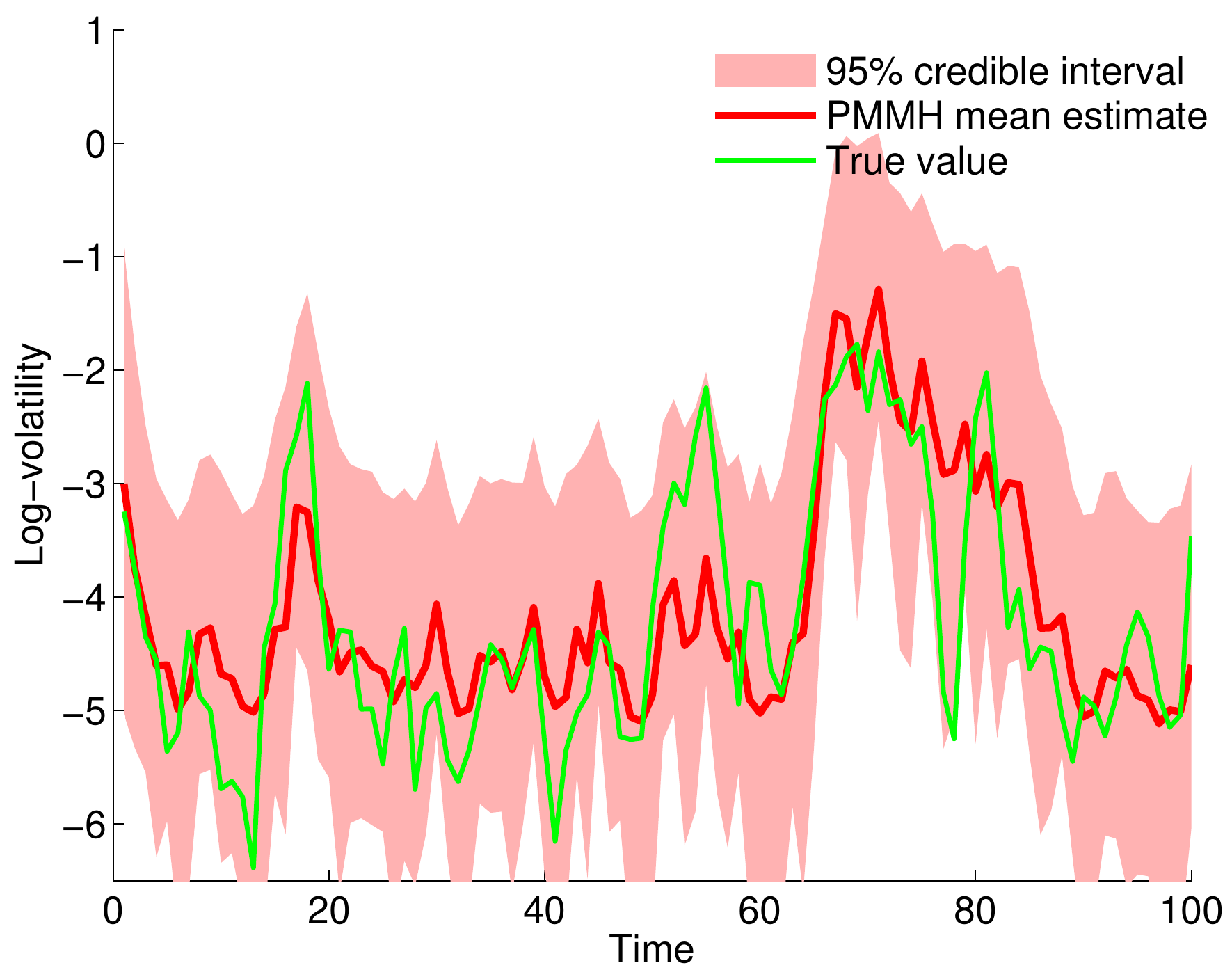}}
\subfigure[Posterior marginals]{\includegraphics[width=.48\textwidth]{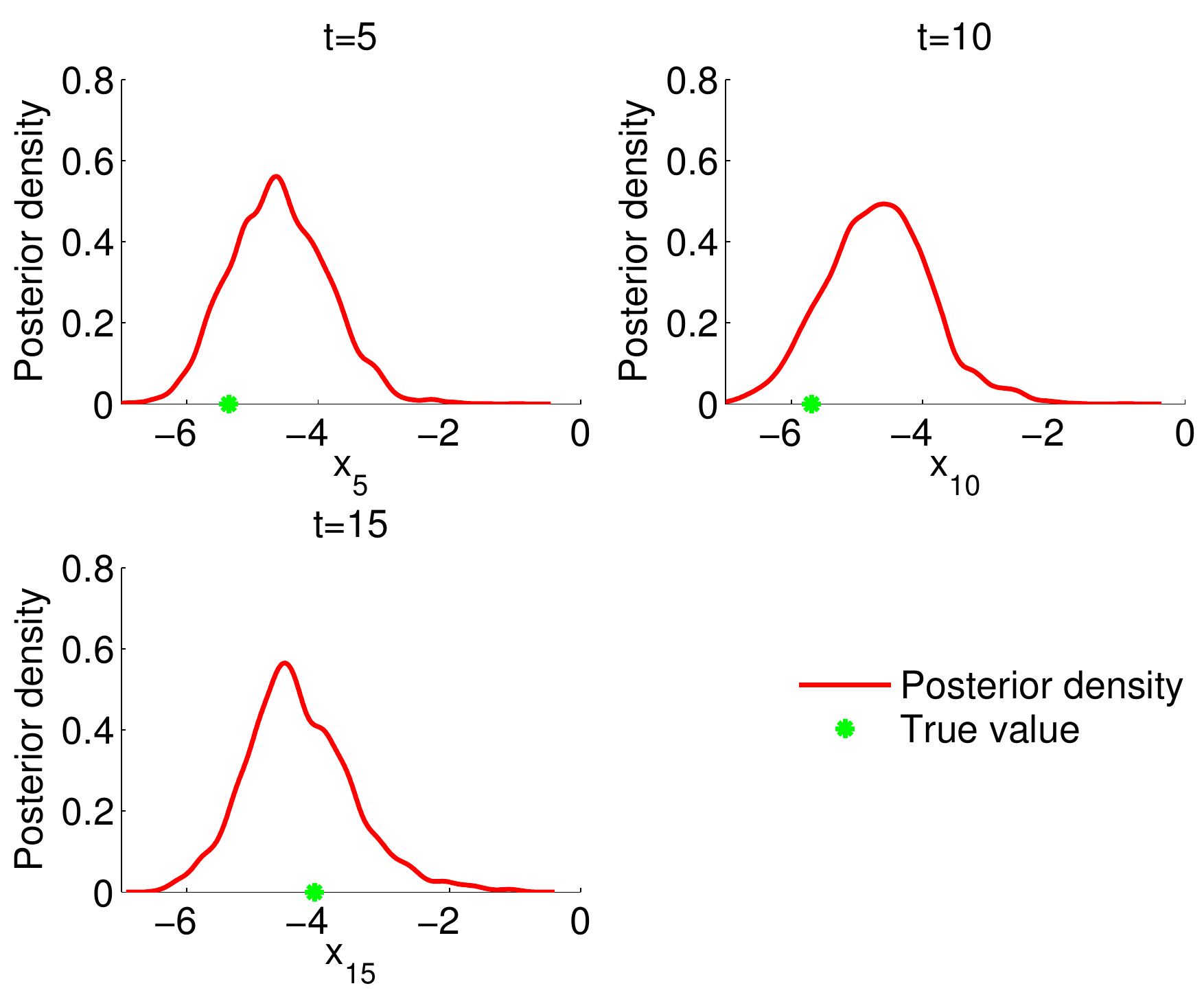}}
\end{center}
\caption{PMMH: (a) Smoothing estimates and credible intervals for the switching stochastic volatility model. (b) Posterior marginals $p(x_t|y_{1:t_{\max}})$ for $t=5,10,15$.}
\label{fig:volatility_pmmh_x}
\end{figure}

\begin{figure}[ht!]
\begin{center}
\includegraphics[width=.5\textwidth]{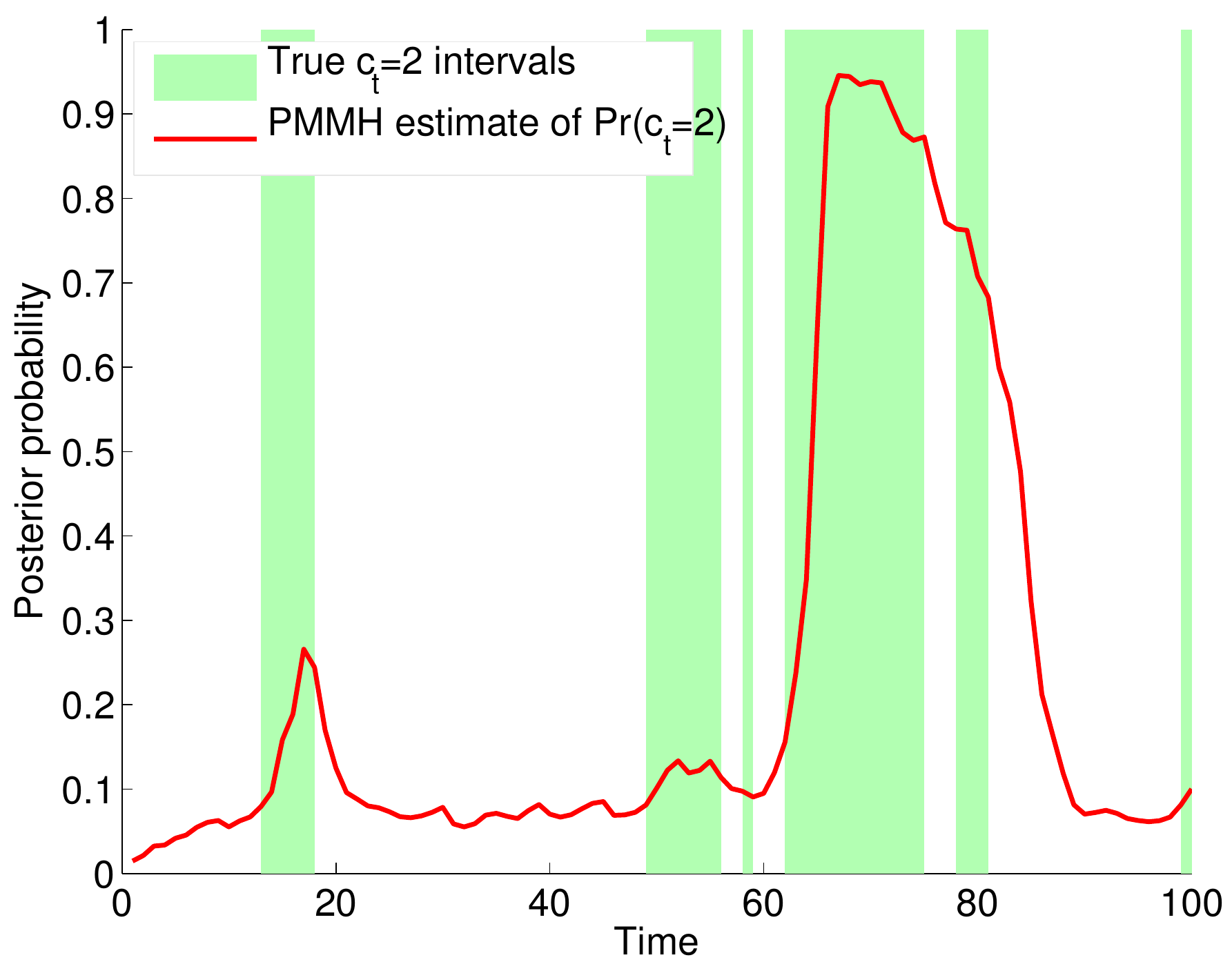}
\end{center}
\caption{PMMH: Marginal posterior probability $\Pr(C_t=2|y_1,\ldots,y_{t_{max}})$ of being in state 2 over time. Shaded green area corresponds to the true state being in state 2.}
\label{fig:volatility_pmmh_c}
\end{figure}

\begin{figure}[ht!]
\begin{center}
\includegraphics[width=.5\textwidth]{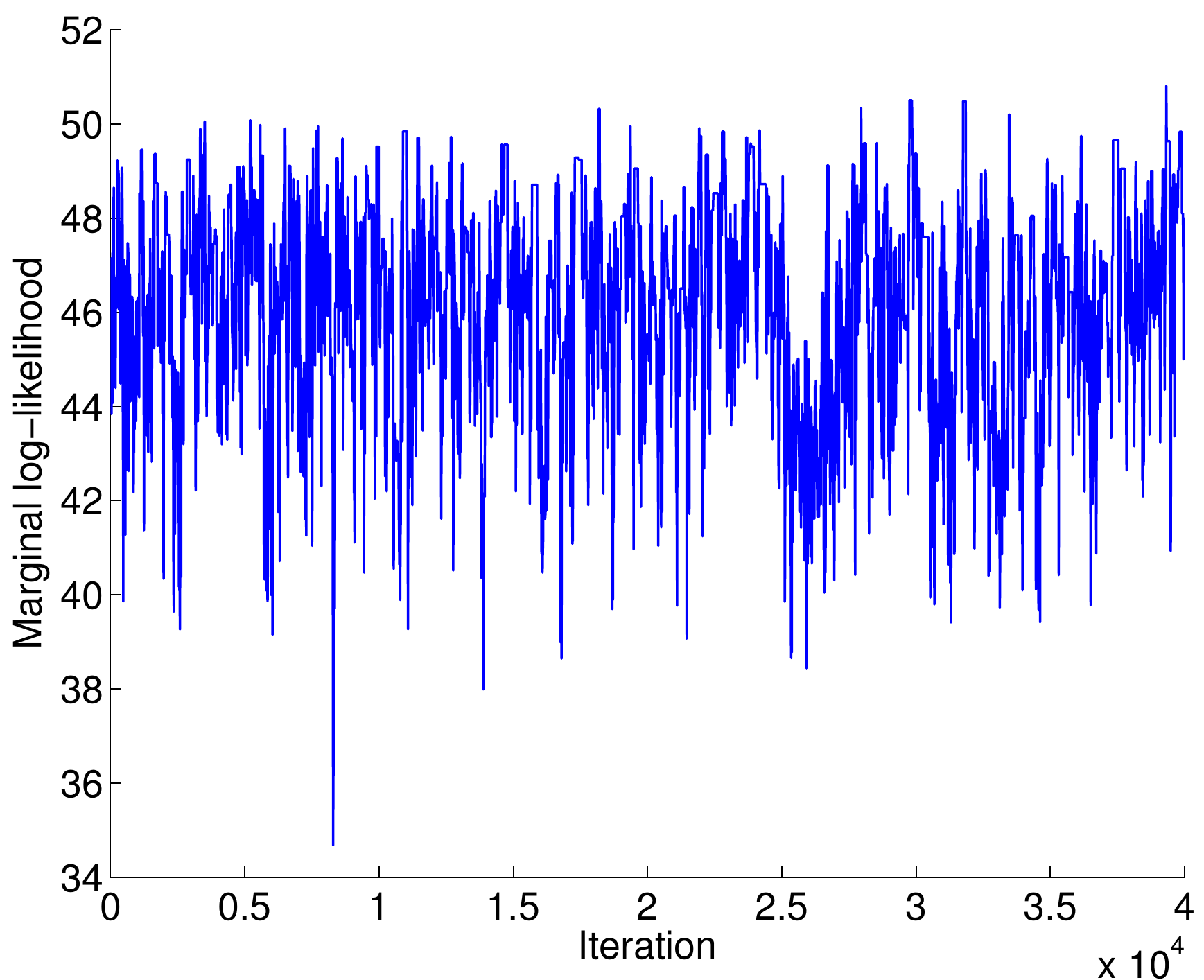}
\end{center}
\caption{PMMH: Logarithm of the marginal likelihood $p(y_1,\ldots,y_{t_{max}}|\theta)$ over MCMC iterations.}
\label{fig:volatility_pmmh_mll}
\end{figure}

Posterior sample traces and histograms of the parameters are given in Figure~\ref{fig:volatility_pmmh_param}. Posterior means and marginal posteriors for the variables $(X_1,\ldots,X_{t_{max}})$ and $(C_1,\ldots,C_{max})$ are given in Figures~\ref{fig:volatility_pmmh_x} and~\ref{fig:volatility_pmmh_c}. The PMMH also returns an estimate of the logarithm of the marginal likelihood at each iteration. The algorithm can therefore also be used as a stochastic search algorithm for finding the marginal MAP of the parameters. The log-marginal likelihood is shown on Figure~\ref{fig:volatility_pmmh_mll}.

\newpage
\section{Example: Stochastic kinetic Lotka-Volterra model}
\label{sec:kinetic}

We consider now Bayesian inference in the Lotka-Volterra model~\citep{Boys2008}. This continuous-time Markov jump process describes the evolution of two species $X_{1}(t)$ (prey) and $X_{2}(t)$ at time $t$, evolving according to the three reaction equations:

\begin{equation}
\begin{tabular}{rcll}
  $X_1$ & $\overset{c_1}{\longrightarrow}$ & $2 X_1$ & ~~prey reproduction, \\
  $X_1+X_2$ & $\overset{c_2}{\longrightarrow}$ &$2X_2$ &  ~~predator reproduction,\\
  $X_2$ & $\overset{c_3}{\longrightarrow}$& $\emptyset $ & ~~predator death
\end{tabular}
\end{equation}
where $c_1=0.5$, $c_2=0.0025$ and $c_3=0.3$ are the rate at which some reaction occur. Let $dt$ be an infinitesimal interval. More precisely, the process evolves as
 \begin{subequations}
\begin{align}
\Pr(X_1(t+dt)=x_1(t)+1,X_2(t+dt)=x_2(t)|x_1(t),x_2(t))&=c_1 x_1(t)dt+o(dt)\\
\Pr(X_1(t+dt)=x_1(t)-1,X_2(t+dt)=x_2(t)+1|x_1(t),x_2(t))&=c_2 x_1(t)x_2(t)dt+o(dt)\\
\Pr(X_1(t+dt)=x_1(t),X_2(t+dt)=x_2(t)-1|x_1(t),x_2(t))&=c_3 x_2(t)dt+o(dt).
\end{align}
\label{eq:kinetic_state}
\end{subequations}

Forward simulation from the model~\eqref{eq:kinetic_state} can be done using the Gillespie algorithm~\citep{Gillespie1977,Golightly2013}. We additionally assume that we observe at some time $t=1,2,\ldots,t_{\max}$ the number of preys with some additive noise
\begin{equation}
Y(t)=X_1(t) + \epsilon(t), ~~\epsilon(t)\sim\Norm(0,\sigma^2)
\label{eq:kinetic_meas}
\end{equation}

The objective is to approximate the posterior distribution on the number of preys and predators $(X_1(t),X_2(t))$ at $t=1,\ldots,t_{\max}$ given the data $(Y(1),\ldots,Y(t_{\max}))$. Listing~\ref{listing:skm} gives the transcription of the model defined by Equations~\eqref{eq:kinetic_state} and~\eqref{eq:kinetic_meas} in the \BUGS\ language.
\begin{lstlisting}[style=bugs,caption=Stochastic kinetic model in BUGS language,label=listing:skm,float]
model
{
  x[,1] ~ LV(x_init, c[1], c[2], c[3], 1)
  y[1] ~ dnorm(x[1,1], 1/sigma^2)
  for (t in 2:t_max)
  {
    x[,t] ~ LV(x[,t-1], c[1], c[2], c[3], 1)
    y[t] ~ dnorm(x[1,t], 1/sigma^2)
  }
}
\end{lstlisting}

The Gillespie sampler to sample from~\eqref{eq:kinetic_state} is not part of the \BUGS\ library of samplers. Nonetheless, \Biips\ allows the user to add two sorts of external functions:
\begin{enumerate}
\item Deterministic functions, with \code{biips_add_function}. Such an external function is called after the symbol \code{<-} in \BUGS, e.g. \code{y <- f_ext_det(x)}
\item Sampling distributions, with the function \code{biips_add_distribution}. Such an external sampler is called after the symbol \code{~} in \BUGS, e.g. \code{z ~ f_ext_samp(x)}\footnote{Note that in the current version of \Biips\, the variable \code{z} needs to be unobserved in order to use a custom distribution.}
\end{enumerate}

 The function \code{`LV'} used in the Listing \ref{listing:skm} is an additional sampler calling a \Matlab/\R\ custom function to sample from the Lotka-Volterra model using the Gillepsie algorithm.

\begin{lstlisting}[style=matbiips]
function x = lotka_volterra_gillespie(x, c1, c2, c3, dt)
% Matlab function to sample from a Lotka-Volterra model
% with the Gillepsie algorithm
z = [1, -1, 0; 0, 1, -1];
t = 0;
while true
  rate = [c1 * x(1), c2 * x(1) * x(2), c3 * x(2)];
  sum_rate = sum(rate);
  % Sample the next event from an exponential distribution
  t = t - log(rand) / sum_rate;
  % Sample the type of event
  ind = find((sum_rate * rand) <= cumsum(rate), 1);
  if t > dt
    break
  end
  x = x + z(:,ind);
end

\end{lstlisting}

\begin{lstlisting}[style=rbiips]
lotka_volterra_gillespie <- function(x, c1, c2, c3, dt) {
  # R function to sample from a Lotka-Volterra model
  # with the Gillepsie algorithm
  z <- matrix(c(1, -1, 0, 0, 1, -1), nrow=2, byrow=TRUE)
  t <- 0
  while (TRUE) {
    rate <- c(c1*x[1], c2*x[1]*x[2], c3*x[2])
    sum_rate <- sum(rate);
    # Sample the next event from an exponential distribution
    t <- t - log(runif(1))/sum_rate
    # Sample the type of event
    ind <- which((sum_rate*runif(1)) <= cumsum(rate))[1]
    if (t>dt)
      break
    x <- x + z[,ind]
  }
  return(x)
}
\end{lstlisting}

\begin{figure}[ht!]
\begin{center}
\subfigure[Ground truth and data]{\includegraphics[width=.48\textwidth]{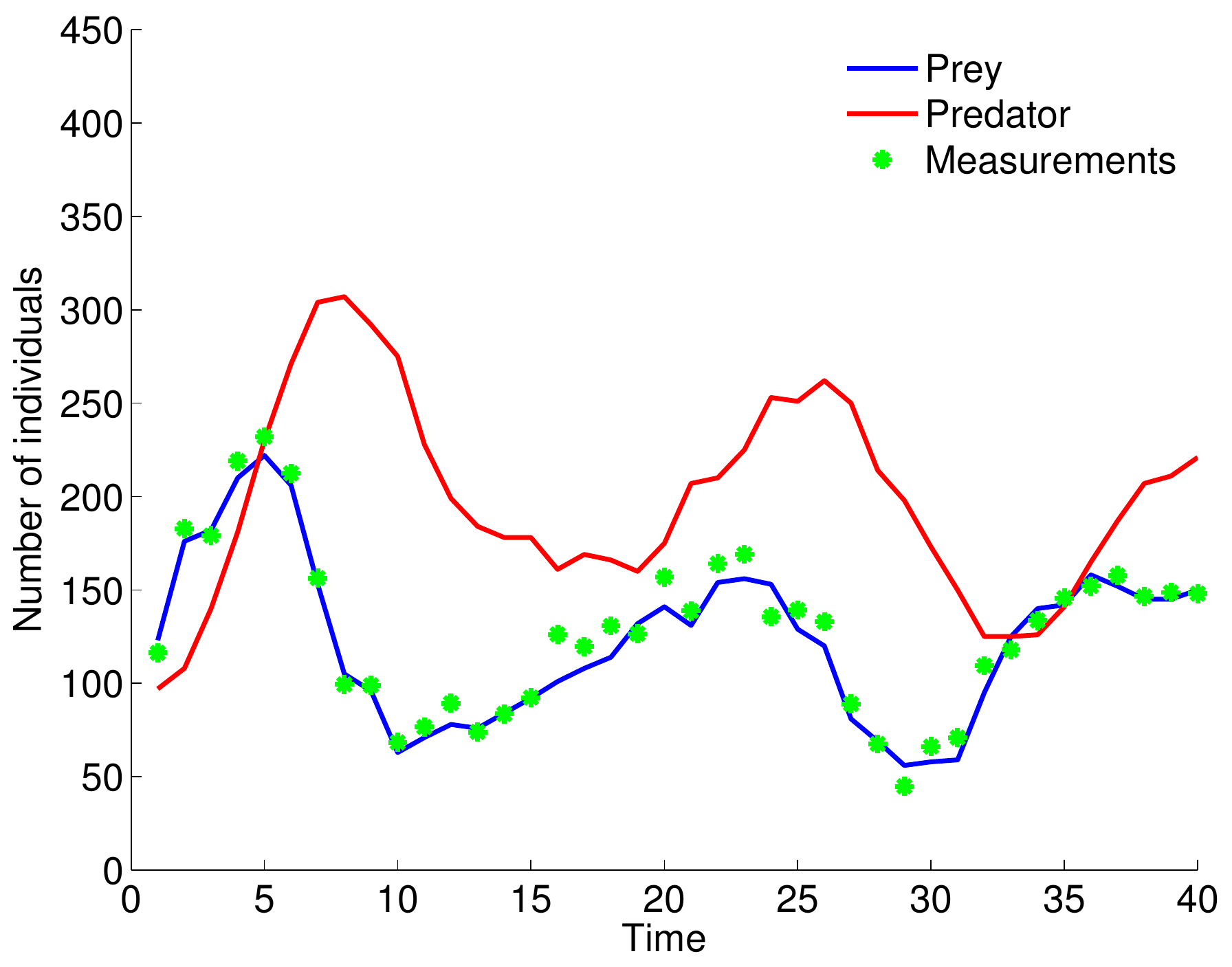}}
\subfigure[Estimates]{\includegraphics[width=.48\textwidth]{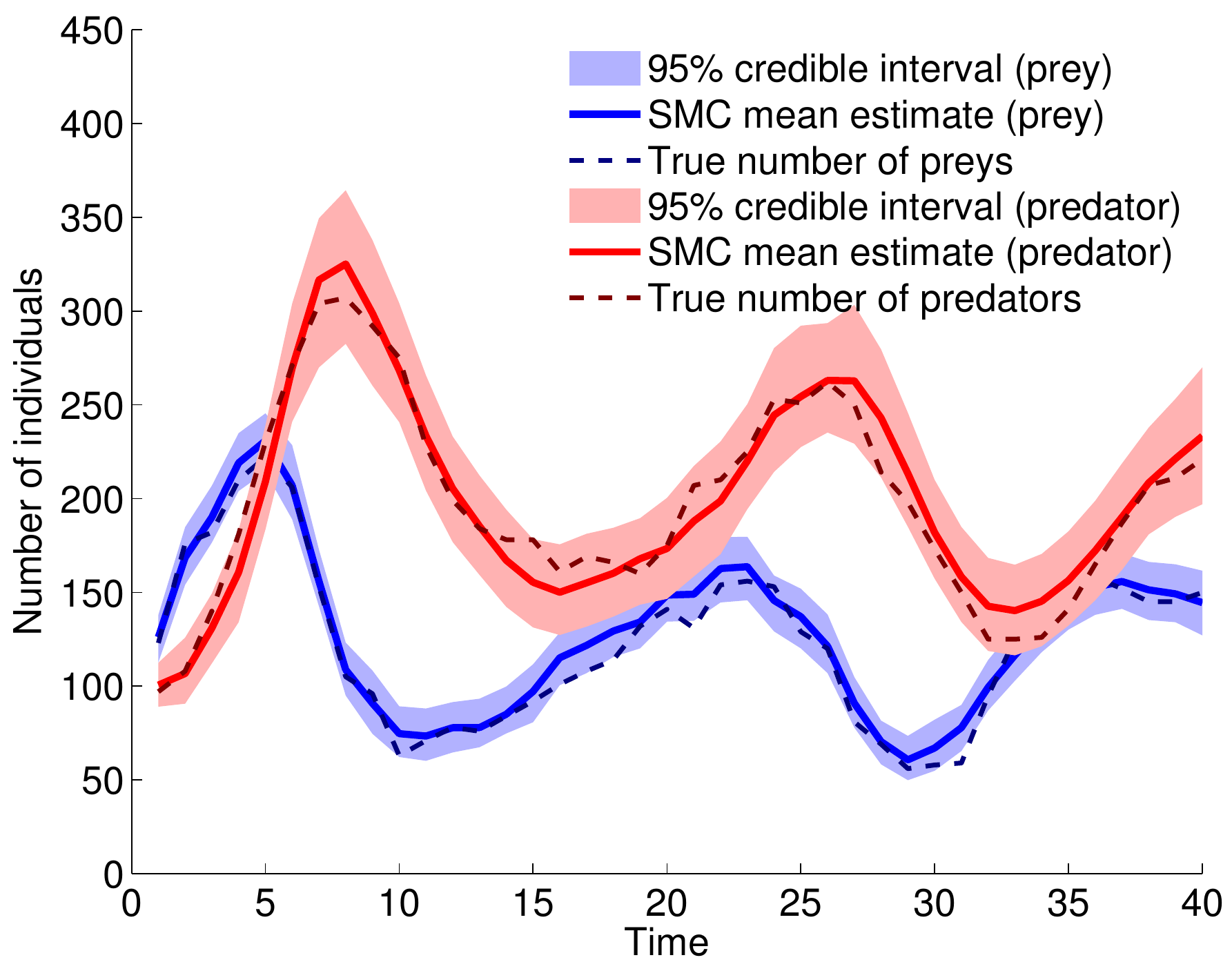}}\\
\subfigure[Smoothing effective sample size]{\includegraphics[width=.48\textwidth]{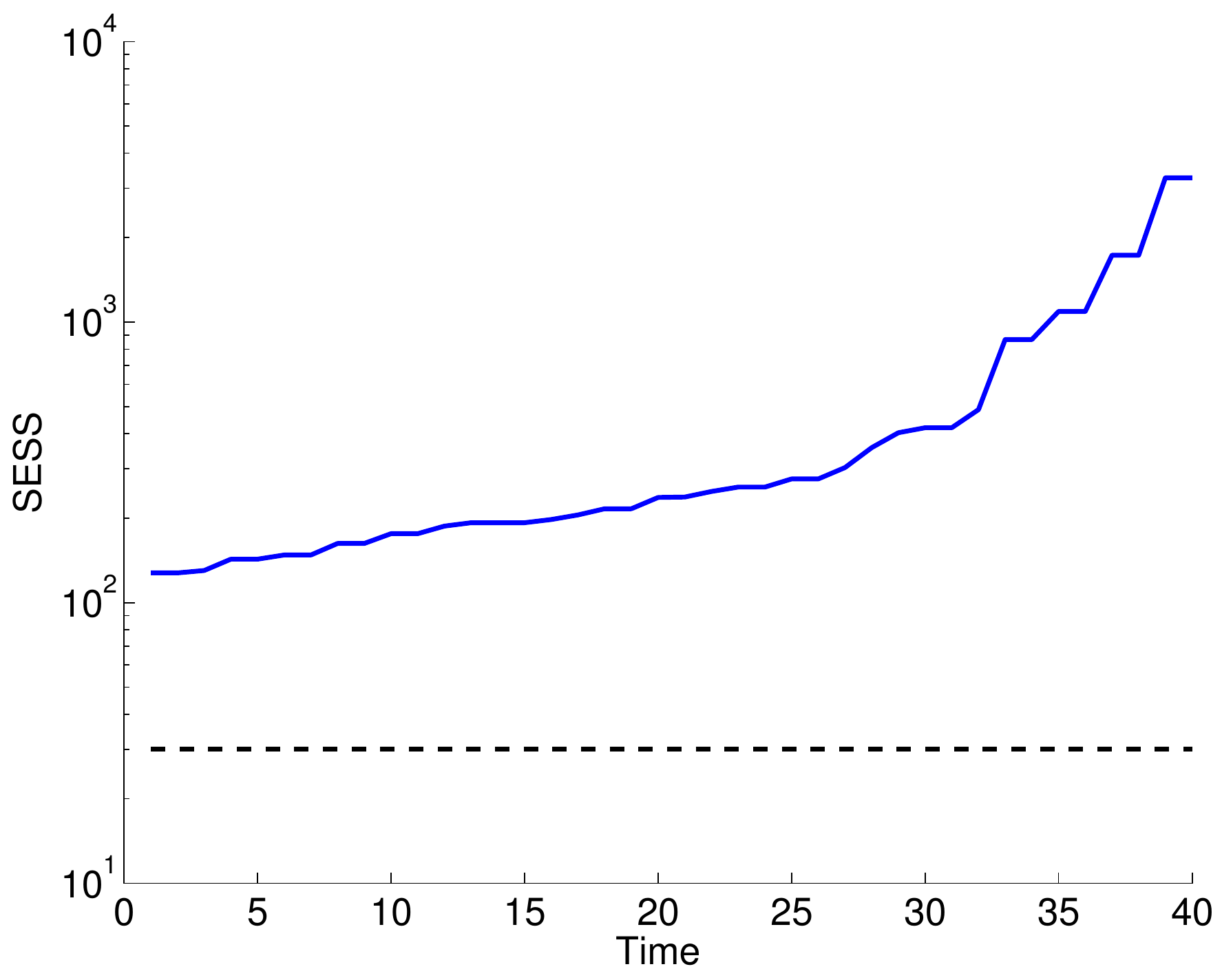}}
\end{center}
\caption{Stochastic kinetic model: (a) True number of prey/predators and measurements. (b) Estimated number of prey/predators and 95\% credible intervals. (c) Smoothing effective sample size.}
\label{fig:kinetic}
\end{figure}

One can add the custom function \code{`LV'} to \Biips, and run a SMC algorithm on the stochastic kinetic model in order to estimate the number of preys and predators. Estimates, together with the true numbers of prey and predators, are reported in Figure~\ref{fig:kinetic}.

\begin{lstlisting}[style=matbiips]
fun_bugs = 'LV'; fun_nb_inputs = 5;
fun_dim = 'lotka_volterra_dim'; fun_sample = 'lotka_volterra_gillespie';
biips_add_distribution(fun_bugs, fun_nb_inputs, fun_dim, fun_sample);

t_max = 40; x_init = [100; 100]; c = [.5, .0025, .3]; sigma = 10;
data = struct('t_max', t_max, 'c', c, 'x_init', x_init, 'sigma', sigma);
model_file = 'stoch_kinetic_gill.bug'; sample_data = true;
model = biips_model(model_file, data, 'sample_data', sample_data);

n_part = 10000; variables = {'x'};
out_smc = biips_smc_samples(model, variables, n_part);

summ_smc = biips_summary(out_smc, 'probs', [.025, .975]);
\end{lstlisting}

\begin{lstlisting}[style=rbiips]
fun_bugs <- 'LV'; fun_nb_inputs <- 5
fun_dim <- lotka_volterra_dim; fun_sample <- lotka_volterra_gillespie
biips_add_distribution(fun_bugs, fun_nb_inputs, fun_dim, fun_sample)

t_max <- 40; x_init <- c(100, 100); c <- c(.5, .0025, .3); sigma <- 10
data <- list(t_max=t_max, c=c, x_init=x_init, sigma=sigma)
model_file <- 'stoch_kinetic_gill.bug'; sample_data=TRUE
model <- biips_model(model_file, data, sample_data=sample_data)

n_part <- 10000 ; variables <- c('x')
out_smc <- biips_smc_samples(model, variables, n_part)

summ_smc <- biips_summary(out_smc, probs=c(.025, .975))
\end{lstlisting}

%
%
%
%
%
%
%
%
%

\section{Discussion of related software}
\label{sec:discussion}

\subsection{Related software for Bayesian inference using MCMC}
\Biips\ belongs to the \BUGS\ language software family of \WinBUGS, \OpenBUGS~\citep{Lunn2000,Lunn2012} and \JAGS\ software~\citep{Plummer2003}. All these probabilistic programming software use \BUGS\ as a language for describing the statistical model.

In particular, \Biips\ is written in \Cpp\ like \JAGS\ but unlike \pkg{Win}/\OpenBUGS\ which is written in \proglang{Component Pascal}. This was a good reason for adapting \JAGS\ implementation of the \BUGS\ language which might slightly differ from the \pkg{Win}/\OpenBUGS\ original one.

Like the above software, \Biips\ compiles the model at runtime, by dynamically allocating instances of node classes. The resulting graphical model might have a substantial memory size. \Stan~\citep{Stan2013} is a similar software which uses another strategy. It translates the model description into \Cpp\ code that is transformed into an executable at compile-time. This might result in lower memory occupancy and faster execution, at the cost of a longer compilation. In addition, \Stan\ implements its own language for model definition. In particular, \Stan\ language is imperative as opposed to the declarative nature of \BUGS.

The main difference between \Biips\ and the aforementioned software is that \Biips\ uses SMC instead of MCMC as inference algorithm.

\subsection{Related software for Bayesian inference using SMC}

\SMCTC~\citep{Johansen2009} is a \Cpp\ template class library offering a generic framework for implementing SMC methods. It does not come with many features though and might require a lot of coding and understanding of SMC from the user. The development of \Biips\ has started by adapting this library and providing it with more user-friendly features. This template is extended in \pkg{vSMC}~\citep{Zhou2013} to directly support parallelisation. \pkg{RCppSMC}~\citep{Eddelbuettel2014} provides an \proglang{R} interface to \SMCTC.

\LibBi~\citep{Murray2013} is another similar software that implements SMC methods and is suited to parallel and distributed computer hardware such as multi-core CPUs, GPUs and clusters. \LibBi\ comes with its own modeling language although restricting to the state-space model framework. Like \Stan, \LibBi\ transforms the model definition into an executable at compile-time, resulting in high computing performances.

Other software such as \pkg{Venture}~\citep{Mansinghka2014} or \pkg{Anglican}~\citep{Wood2014} also propose particle MCMC inference engines for posterior inference, with a different probabilistic language, that allows for more expressiveness; in particular, it can deal with models of changing dimensions, complex control flow or stochastic recursion.

\subsection{Related software for stochastic optimization}

Although the focus of \Biips\ is automatic Bayesian inference, interacting particle methods have long been successfully used for stochastic optimization. These algorithms use similar exploration/selection steps and are usually known under the names of evolutionary algorithms, genetic algorithms or meta-heuristics. Several softwares have been developed over the past few years, such as \pkg{EASEA}~\citep{Collet2000}, \pkg{Evolver}\footnote{http://www.palisade.com/evolver/} or \pkg{ParadisEO}~\citep{Cahon2004}.

\section{Conclusion}
\label{sec:conclusion}

The \Biips\ software is a \BUGS\ compatible, black-box inference engine using sequential Monte Carlo methods. Due to its use of the \BUGS\ language, and the ability to define custom functions/distributions, it allows a lot of flexibility in the development of statistical models. By using particle methods, the software can return estimates of the marginal likelihood at no additional cost, and can use custom conditional distributions, possibly with an intractable expression. Although particle methods are particularly suited to posterior inference on the two examples discussed in this paper, \Biips\ running times are still higher than those of a more mature software with a MCMC inference engine such as \JAGS. Nonetheless, there is room for improvement and optimization of \Biips; in particular, particle algorithms are particularly suited to parallelization~\citep{Lee2010,Verge2013,Murray2013}, and we plan in future releases of the software to provide a parallel implementation of \Biips .

\paragraph{Acknowledgement.} The authors thank Arnaud Doucet, Pierre Jacob, Adam Johansen and Frank Wood for useful feedback on earlier versions of the paper. Fran\c cois Caron acknowledges the support of the European Commission under the Marie Curie Intra-European Fellowship Programme.

\bibliographystyle{jss}
\bibliography{biipsbib}

\end{document}